\documentclass[preprint,11pt,authoryear]{elsarticle}

\usepackage{amsmath,amssymb}
\usepackage{array}
\usepackage{booktabs}
\usepackage{multirow}
\usepackage{makecell}
\usepackage{graphicx}
\usepackage{rotating}
\usepackage{placeins}
\usepackage{colortbl}
\usepackage{xcolor}
\usepackage{textcomp}
\usepackage{enumitem}
\usepackage{needspace}
\usepackage{eurosym}
\usepackage{xurl}
\usepackage{subcaption}
\usepackage{etoolbox}
\usepackage{mdframed}
\usepackage{hyperref}
\hypersetup{pdfauthor={Sahab Zandi, Noah Kostesku, Christophe Mues, Maria Oskarsdottir, Cristian Bravo},breaklinks=true}
 
\graphicspath{{./images/}}
\usepackage[margin={1in,1in}]{geometry}
\newcolumntype{L}[1]{>{\raggedright\arraybackslash}p{#1}}

\newmdenv[
  linecolor=black,
  linewidth=0.5pt,
  backgroundcolor=gray!5,
  roundcorner=2pt,
  innertopmargin=4pt,
  innerbottommargin=4pt,
  innerleftmargin=6pt,
  innerrightmargin=6pt
]{exampleframe}

\usepackage{multirow}
\usepackage{setspace}
\usepackage{natbib}
\usepackage{booktabs}
\usepackage{makecell}
\usepackage{caption}
\usepackage{amsmath}
\usepackage{tabularx}
\usepackage{longtable}
\usepackage{enumitem}
\usepackage{threeparttable}
\usepackage{array}
\newcolumntype{L}[1]{>{\raggedright\arraybackslash}p{#1}}
\usepackage{mdframed}
\usepackage{needspace}

\usepackage{siunitx}
\journal{European Journal of Operational Research}

\begin{document}

\begin{frontmatter}
\title{Communicating Credit Risk with Large Language Models: Evaluation of Explanations from Standard and Alternative Data-Based Models}

 \author[inst1]{Sahab Zandi}
 \author[inst2]{Noah Kostesku}
 \author[inst3,inst4]{Christophe Mues}
 \author[inst5,inst6]{Mar\'ia \'Oskarsd\'ottir\corref{cor1}}
 \ead{m.oskarsdottir@soton.ac.uk}
 \cortext[cor1]{Corresponding author.}
 \author[inst1]{Cristi\'an Bravo}

 \affiliation[inst1]{organization={Department of Statistical and Actuarial Sciences, Western University},
             city={London},
             country={Canada}}
 \affiliation[inst2]{organization={Department of Computer Science, Western University},
             city={London},
             country={Canada}}
 \affiliation[inst3]{organization={University of Southampton Business School, University of Southampton},
             city={Southampton},
             country={United Kingdom}}
 \affiliation[inst4]{organization={Centre for Operational Research, Management Sciences and Information Systems, University of Southampton},
             city={Southampton},
             country={United Kingdom}}
\affiliation[inst5]{organization={School of Mathematical Sciences, University of Southampton},
             city={Southampton},
             country={United Kingdom}}
 \affiliation[inst6]{organization={Department of Computer Science, Reykjavik University},city={Reykjavik},
             country={Iceland}}

\begin{abstract}
Credit decisioning is a high-stakes task in which model outputs must be accurate and explainable to support compliant decisions. Although modern credit risk models such as eXtreme Gradient Boosting (XGBoost) and Graph Neural Networks (GNNs) improve predictive performance, their explanations are often too technical for stakeholders, such as auditors and customers, creating communication gaps that can shape approvals, denials, and fairness judgments. We examine whether Large Language Models (LLMs) can serve as explanation layers that translate post-hoc explanation artefacts into stakeholder-appropriate risk narratives. 
Using Freddie Mac single-family loan-level data, we develop three pipelines: standard tabular (XGBoost + SHAP), and two models with alternative data, a pure network-based model (GNN + GNNExplainer), and a bimodal model (combining tabular and network data). We generate narratives with three LLM configurations: a small fine-tuned LLM (Gemma~3~4B), a large fine-tuned LLM (DeepSeek~R1~70B), and a zero-shot commercial LLM (Gemini~2.5). Explanation quality is evaluated through automated checks across all pipelines and a human study of bimodal explanations comparing credit risk professionals and non-professionals on eight decision-relevant dimensions.
We have three main findings. First, the pipeline accounts for higher variance in evidence-grounding scores than the language model, meaning that the binding constraint on explanation quality is the evidence representation, not the model used. Second, the explanation narratives reliably name the influential factors but are less reliable when stating the direction of influence, which may be consequential for adverse-action communication. Finally,  professionals apply stricter evidentiary standards than non-professionals. We discuss implications for the governance of risk models, including deployment considerations and the value of domain-aligned LLMs in regulated credit settings.
\end{abstract}


\end{frontmatter}

\section{Introduction}
\label{Introduction}

Credit decisioning is a high-stakes risk assessment: mortgage and consumer lending decisions allocate capital under uncertainty and can materially affect household welfare, institutional losses, and perceptions of fairness \citep{doumpos2023operational,tigges2024gets}. Predictive accuracy alone is therefore insufficient; institutions must also justify model-driven decisions in ways that support oversight, customer communication, and internal governance \citep{de2024explainable}. Regulatory and policy frameworks reinforce this requirement through Basel-style risk governance, data protection rules such as GDPR in the EU, and adverse-action communication requirements in the U.S. \citep{bis2011basel3,cfpb2022circular,european2016gdpr}. As a result, the quality of the explanation becomes part of the risk itself: poorly grounded or poorly communicated explanations can contribute to miscalibrated trust, inconsistent approvals/denials, operational disputes between the validation and business teams, and compliance exposure \citep{fischhoff1995risk}.

At the same time, credit risk modelling has moved toward a broader set of complex machine learning methods that can improve predictive performance but reduce transparency. These include tree-based ensembles and deep learning models \citep{hayashi2022dlcreditscoring,shi2022mlcreditrisk}. Recent research also suggests that credit risk may exhibit a relational structure \citep{oskarsdottir2021multilayer}. Borrowers linked by shared contextual factors, such as geography or lender practices, can exhibit correlated risk patterns that traditional tabular models fail to capture. This has motivated network-based approaches, especially GNNs, for credit risk modelling \citep{zandi2025attention}.

Post-hoc explanation methods help surface model drivers, for example, SHAP for tabular models \citep{lundberg2017shap} and GNNExplainer for GNNs \citep{ying2019gnnexplainer}. However, their outputs are typically technical artefacts (feature attributions, node/edge importance scores, local subgraphs) that may not translate into the concise, audience-appropriate narratives needed by stakeholders such as risk professionals, auditors, regulators, and affected customers. This creates a persistent risk-communication gap: models may be explainable in principle, yet the explanations are not consistently usable in practice.

LLMs offer a potential bridge. By converting structured input into coherent natural-language narratives, LLMs can act as a layer that translates the explanation output into stakeholder-appropriate text. However, in regulated environments, the central issue is not only understandability but also whether narratives remain faithful and relevant to decisions \citep{turpin2023language}.

This paper presents a unified framework that converts explanation artefacts (SHAP and GNNExplainer output) into natural-language narratives across tabular, network, and bimodal evidence. We use the framework to study LLM-based explanation layers for credit risk models through three pipelines that separate prediction, explanation generation, and narrative generation. Using Freddie Mac single-family loan-level data \citep{freddiemac2022}, we develop (i) a tabular-based pipeline (XGBoost + SHAP), (ii) a network-based pipeline (GNN + GNNExplainer) and (iii) a bimodal pipeline that combines both types of evidence. We compare three LLM configurations representing realistic deployment choices: a small domain-fine-tuned open LLM (Gemma~3~4B), a large domain-fine-tuned open LLM (DeepSeek~R1~70B), and a general-purpose zero-shot commercial LLM (Gemini~2.5).

Because explanations in this domain serve different stakeholders, we distinguish professional and non-professional evaluators. Prior work in Explainable Artificial Intelligence (XAI) and risk communication shows that explanation needs vary by audience, and that NCRP readers may respond differently to the frame, tone, and format of the explanation \citep{dieckmann2009use,DoshiVelez2017,miller2019explanation,peters2008numeracy}. Given this, we combine automated checks, which provide scalable guardrails for fluency and evidence alignment, with human ratings, which capture whether explanations are perceived as trustworthy, communicable, and usable in realistic decision settings.

We make three contributions. First, we decompose explanation quality in a credit risk decision-support system into its sources, namely evidence modality, language model, and evaluator, and show that the first dominates the second by roughly an order of magnitude, which relocates the design problem from model selection to evidence representation. Second, we identify a systematic asymmetry between what generated explanations cover and what they get directionally right, and argue that directional error, not omission, is the governance-relevant failure mode in regulated credit communication. Third, we show that professional and non-professional stakeholders apply
different acceptance criteria to the same narrative, and that rater heterogeneity dominates system differences in human evaluation. From this we derive design requirements for evaluating explanation layers, responding to the evaluation agenda set out for explainable AI in operational research by \citet{de2024explainable}.
We address the following research questions.

\begin{itemize}
    \item RQ1: How does explanation modality (tabular, network, bimodal) influence automated measures of explanation quality across LLM configurations?
    \item RQ2: Do LLM-generated explanations differ in perceived explanation quality between professionals and non-professionals?
    \item RQ3: How do the LLM choice and fine-tuning status relate to the quality of the explanation?
    \item RQ4: Which textual properties are associated with a higher human-rated explanation quality?
\end{itemize}

The remainder of the paper proceeds as follows: Section~\ref{Background and Related Work} reviews related work and regulatory context; Section~\ref{Methodology} details the three-stage explanation architecture, data, models, and evaluation design; Section~\ref{Results} reports findings relevant to the research questions. Section~\ref{Discussion} discusses implications for credit risk communication, the mechanisms underlying the observed effects, and practical and policy considerations. Section~\ref{Conclusions} presents conclusions and directions for future research.

\section{Background and Related Work}
\label{Background and Related Work}

\subsection{Explainability in credit risk modelling}
\label{Explainability in credit risk modelling}

Interpretability has long been central in credit risk, where traditional models such as logistic regression and scorecards were favoured in part because they are transparent by design \citep{Siddiqi2020,Thomas1999}. As institutions increasingly adopt higher-capacity machine learning models, predictive performance can improve, but transparency often decreases \citep{Addo2018,Breeden2020, bucker2022transparency,Lessmann2015}. Post-hoc explainers such as SHAP and LIME have therefore become common tools for summarising feature influence in complex models, including credit-scoring applications \citep{Bracke2019,Bussmann2020,lundberg2017shap,Misheva2021,ribeiro2016should}. 

Importantly, there has been a call for stakeholder-appropriate explanations and evaluation standards \citep{de2024explainable}, highlighted in part by explanation stability of both SHAP and LIME which both cost-sensitive learning and class imbalance can affect \citep{ballegeer2025evaluating,chen2024interpretable}.

However, multiple surveys and conceptual frameworks emphasise that the quality of the explanation depends on the stakeholder and that technical transparency does not guarantee the usefulness of communication \citep{DoshiVelez2017,Guidotti2018}. In regulated settings, a further concern is that post-hoc explanations can be misleading if they are interpreted causally or if they omit important context. Recent calls for inherently interpretable models highlight these concerns and argue that some domains may require models that are transparent by design \citep{rudin2019stop,Rudin2019, tu2025inherently}.

For the purposes of this paper, we treat SHAP-style feature attributions as widely adopted, theoretically grounded explanation artefacts for tabular models. Our focus is not to propose new attribution methods, but to evaluate how such artefacts can be translated into stakeholder-appropriate narratives without sacrificing evidence grounding.

\subsection{Graph-based credit risk modelling and explainability}
\label{Graph-based credit risk modelling and explainability}
Relational modelling has gained traction in credit risk because borrower outcomes can exhibit dependence structures linked to shared geography, lender practices, and other common contextual attributes. Network-based approaches allow such relationships to be explicitly represented and have been shown to improve credit risk modelling performance in settings where borrowers are connected through common contextual factors \citep{oskarsdottir2021multilayer}. GNNs provide a flexible framework to take advantage of such relational structures \citep{wu2020comprehensive}, and recent studies report predictive gains when relational information is incorporated into credit risk models \citep{das2023credit,liu2024unveiling,wang2021temporal,Wang2022,zandi2025attention}. However, interpretability becomes more complex: predictions may depend on both node attributes (borrower/loan features) and graph structure (neighbourhood composition, cluster effects, and message passing).

Alternative GNN explainers include surrogate- and counterfactual-based methods \citep{huang2023graphlime, lucic2022cfgnnexplainer}; we use GNNExplainer as a widely adopted instance-level baseline \citep{kakkad2023surveygnn}. Converting graph explanations into narratives carries a risk of drifting beyond what structural evidence supports; our approach therefore emphasises structured evidence blocks and explicit constraints to reduce over-interpretation.

\subsection{Risk communication, governance, and LLM-based explanation layers}
\label{subsec:risk-comm-governance-llm}
Narrative explanations can improve stakeholder understanding relative to raw attribution output \citep{martens2023tell}, but persuasive language can outpace factual support, motivating evaluation of both readability and evidence grounding. Modern credit models are embedded in governance processes where explanations serve as controls \citep{de2017proposed,souza2023ai}: when an LLM delivers explanations, prompting rules, output constraints, and versioning can materially affect downstream interpretations even when the predictive model is unchanged. A fluent but weakly grounded explanation increases operational and compliance risk by creating a false sense of evidentiary adequacy. Together, Basel-style governance, GDPR, and U.S.\ adverse action requirements motivate evaluation dimensions beyond fluency: fidelity and trustworthiness as proxies for audit readiness, and communicability and usability as proxies for operational and customer-facing constraints \citep{bis2011basel3, cfpb2022circular,european2016gdpr}

LLMs can verbalise technical artefacts like feature attributions, rule lists, and counterfactuals into audience-appropriate narratives \citep{ bilal2025llms,brown2020language, martens2023tell}, which is especially attractive in regulated credit settings where explanations must support review, validation, and compliance documentation. However, the same flexibility creates risk: generated narratives can be fluent and persuasive even when they overstate, omit, or mischaracterise the underlying evidence \citep{turpin2023language}. In regulated settings, this makes fidelity, constraint design, and stakeholder-appropriate usefulness the central criteria for evaluating LLM-mediated explanations \citep{turpin2023language}. Explanation quality is also context-dependent: professionals prioritise evidentiary adequacy and traceability whereas non-professionals weight clarity and completeness \citep{DoshiVelez2017, miller2019explanation}, motivating cohort-based evaluation linked to decision relevance and governance use cases. Building on these insights, we pair cohort-based human evaluation with automated checks of evidence grounding to assess whether explanation layers improve decision-relevant understanding while preserving fidelity and auditability across realistic deployment choices.

\section{Methodology}
\label{Methodology}

This section outlines the framework used to develop and evaluate our explanation pipelines. We describe data processing, network construction, predictive models, post-hoc explainers, LLM-based narration architecture, prompt design, fine-tuning, and human, automated, and linguistic evaluation procedures.

\subsection{Overview of explanation architecture}
\label{Overview of explanation architecture}
Our framework is organised as separable stages (prediction, explanation generation, and narrative generation) to support the evaluation of how explanation artefacts and LLM configuration affect perceived explanation quality. In all pipelines, the LLM is used as a verbalisation layer: it receives structured evidence blocks derived from SHAP or GNNExplainer (plus lightweight normalisation and templating) and produces a constrained narrative explanation. The LLM does not alter predictive outputs or raw attribution values.

To make comparisons interpretable, we separate what varies between conditions from what is kept constant. Across all configurations, we hold fixed the loan cases, evidence-block schema, and prompting rules and output constraints (e.g., requirements to reference only evidence-block content and avoid naming explanation algorithms in borrower-facing text). What varies is the evidence modality (tabular, network, or bimodal) and the LLM configuration (Gemma 3, DeepSeek R1, Gemini 2.5).

\subsection{Dataset}
\label{Dataset}
We used the single-family loan-level dataset from the Federal Home Loan Mortgage Corporation (FHLMC; Freddie Mac) \citep{freddiemac2022}, focusing on loans originated during 2015--2016, in line with previous credit risk modelling work on the same data source, although for different years \citep{zandi2025attention}. For each origination cohort, performance records are observed up to a fixed cut-off date, and we retain loans with at least six months of servicing history to reduce bias from insufficient outcome observation while preserving adequate sample coverage. To maintain temporal separation, the 2015 origination cohort is used for model development, and the 2016 cohort is held out for evaluation. Standard preprocessing steps include missing-value imputation, categorical recoding, and min--max scaling. 
A summary of the borrower- and loan-level features used is provided in \ref{app:Features}.

\subsection{Network construction}
\label{Network construction}
\sloppy{We construct static networks from origination-stage data aggregated over fixed time windows: 6-month periods (January to June 2015 for training and January to June 2016 for testing) and a 1-month period (July 2016) for explanation generation. Graph construction uses only origination-stage attributes to maintain temporal separation and avoid outcome leakage. Within each period, nodes correspond to individual loans, and edges connect loans that share a zip code prefix (regional grouping), a common lender, or both, forming fully connected bidirectional cliques. Edge types are one-hot encoded to distinguish connection sources. Node labels use the same binary default definition described in the feature table in \ref{app:Features}.

\subsection{Predictive model training}
\label{Predictive model training}
For the tabular-based pipeline, we train an XGBoost classifier using the 2015 origination cohort and evaluate it on the held-out 2016 cohort. The predictive feature set consists of the preprocessed borrower and loan variables from the origination-stage (described in the feature table in \ref{app:Features}, excluding identifiers, temporal metadata and grouping variables not intended for tabular prediction. Hyperparameters are selected by grid search with 5-fold cross-validation (ROC--AUC criterion); the classification threshold is chosen on a validation split by maximising F1.

For the network-based pipeline, we train a Graph Attention Network (GAT) \citep{velivckovic2018graph} on the 2015 training graph and evaluate it on the held-out 2016 test graph. The model operates on node features derived from origination-stage borrower and loan attributes together with edge attributes that encode connection type and group-size information. The architecture uses stacked attention-based message-passing layers with edge-aware transformations, followed by a binary node-level prediction head for default classification. Training uses neighbour sampling; class imbalance is addressed by approximate 1:5 negative subsampling. Model selection uses validation ROC--AUC with early stopping and F1-based threshold selection.

\subsection{Explainability techniques}
\label{Explainability techniques}
We apply two post-hoc explanation methods aligned with our predictive models: SHAP for tree-based models \citep{lundberg2017shap} and GNNExplainer for GNNs \citep{ying2019gnnexplainer}. SHAP produces feature-level attribution scores for XGBoost predictions and is widely used in credit risk modelling due to its theoretical grounding and compatibility with tree-based ensembles. GNNExplainer identifies subgraph structures and node features driving GNN decisions and provides instance-level explanations directly from the trained model. A detailed formulation of both methods is provided in \ref{app:Mathematical definitions of explainability methods}.

\subsection{Large language models}
\label{Large language models}
LLMs in this framework convert structured output from SHAP and GNNExplainer into human-readable narratives. A lightweight post-processing layer contextualises the explainer outputs (e.g., ranking features, mapping values to percentiles, and generating short descriptive phrases), providing the LLM with structured, enriched inputs for narrative generation. We compare three configurations spanning size, fine-tuning status, and deployment mode (Table~\ref{tab:llms}): Gemma 3 4B (small, fine-tuned), DeepSeek R1 70B (large, fine-tuned), and Gemini 2.5 (undisclosed size, zero-shot API).

\begin{table*}[hbt!]
\caption{LLMs used for narrative generation.}
\label{tab:llms}
\small
\centering
\begin{tabular}{lcccc}
\toprule
LLM & Size & Fine-tuned & Deployment Mode & Source\\
\midrule
Gemma~3 & Small (4B) & Yes & Open-source & Google DeepMind\\
DeepSeek~R1 & Large (70B) & Yes & Open-source & DeepSeek AI\\
Gemini~2.5 & Undisclosed & No & API-based & Google\\
\bottomrule
\end{tabular}
\end{table*}

\subsection{Proposed explanation pipelines}
\label{Proposed explanation pipelines}
We develop three explanation pipelines aligned with different evidence modalities (tabular-based, network-based, and bimodal), each composed of separable stages for model training, explanation generation, and LLM-based narration. This separation enables analysis of how explanation strategies influence interpretability between evidence types. The architecture of the three pipelines, shown in Figure~\ref{fig:pipelines}, can be summarised as follows:

\begin{figure*}[hbt!]
\centering
\subfloat[Tabular-based pipeline\label{fig:pipeline-tabular}]{%
  \includegraphics[scale=0.25]{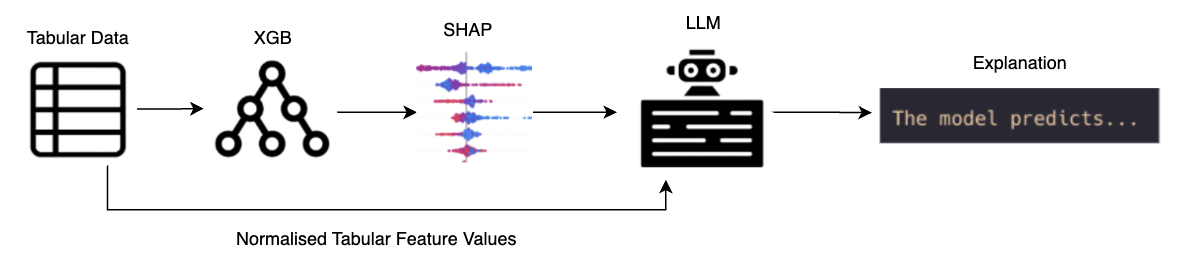}}\\[6pt]
\subfloat[Network-based pipeline\label{fig:pipeline-network}]{%
  \includegraphics[scale=0.25]{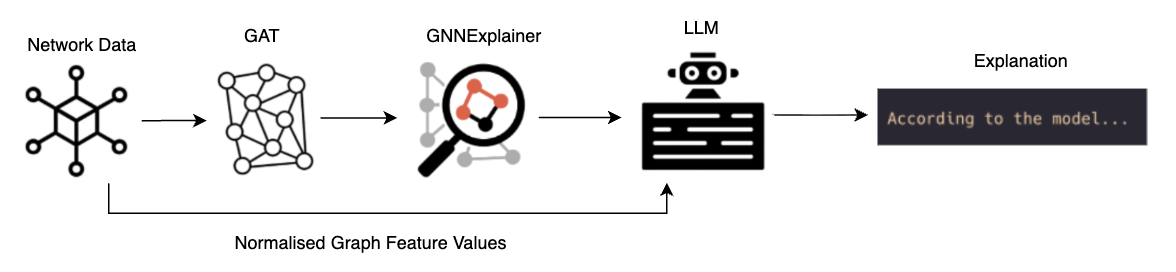}}\\[6pt]
\subfloat[Bimodal pipeline\label{fig:pipeline-bimodal}]{%
  \includegraphics[scale=0.25]{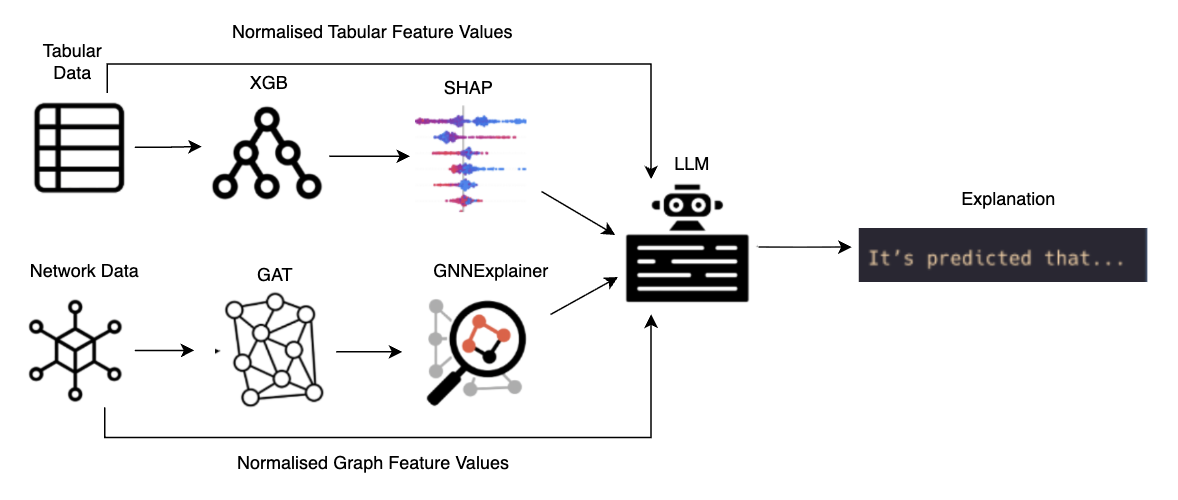}}
\caption{The architecture of the three explanation pipelines.}
\label{fig:pipelines}
\end{figure*}

\begin{itemize}
    \item Tabular-based pipeline: An XGBoost model generates default-risk predictions from loan-level tabular features, and SHAP produces per-instance feature attributions. Before prompting the LLM, a lightweight rule-based post-processing layer converts the top-$k$ SHAP-ranked features into structured textual inputs by mapping selected feature values to hand-crafted plain-language descriptors and qualitative ranges, while preserving the directional contribution indicated by the SHAP scores. The LLM then uses the predicted outcome together with these enriched feature summaries to produce a constrained narrative explanation (Figure~\ref{fig:pipeline-tabular}).
    \item Network-based pipeline: A GAT generates node-level default-risk predictions by combining loan attributes with message passing over the constructed loan graph. GNNExplainer then produces a local explanation consisting of influential node features and a compact explanatory subgraph. The LLM receives a structured summary of this evidence (e.g., salient neighbourhood composition and edge-type signals) and generates a narrative explanation (Figure~\ref{fig:pipeline-network}).
    \item Bimodal pipeline: XGBoost and GAT are run in parallel, and SHAP and GNNExplainer explanations are produced for the same instance. The LLM is provided with both evidence blocks and instructed to synthesise tabular and relational signals into a single narrative while preserving evidence provenance and avoiding unsupported causal claims (Figure~\ref{fig:pipeline-bimodal}).
\end{itemize}

\subsection{Prompt design for explanation generation}
\label{Prompt design for explanation generation}
To transform model evidence into narratives, we design pipeline-specific prompt templates, each combining a shared system prompt with a tailored user prompt. The system prompt enforces objectivity, minimises bias, and establishes stylistic consistency. It avoids persona cues or anthropomorphic framing, which prior work shows can reduce accuracy and increase stylistic drift \citep{gupta2023bias,liu2024evaluating}.

The user prompt adapts to each pipeline using three strategies: (1) instructional step-by-step reasoning \citep{wei2022chain}, (2) task decomposition via explicit subgoals and constrained output structure, and (3) counterfactual-style prompts to elicit improvement-orientated alternatives, drawing on the broader notion of counterfactual reasoning \citep{Pearl2009} and recent work on prompting LLMs for counterfactual generation \citep{li2023counterfactualprompting}. These strategies are realised through explicit prompt components (ordered reasoning steps, guided subgoals, counterfactual prompts, and explicit length/format constraints) rather than free-form generation. Within our experimental setup, Gemma~3 and Gemini~2.5 were prompted using the same prompt templates, while DeepSeek~R1 required minor calibration to account for LLM-specific alignment and formatting differences \citep{deepseek2024}. Table~\ref{tab:prompt-components} summarises the prompt components used in each pipeline. Expanded prompt examples are provided in \ref{app:Prompt examples}.

\begin{table*}[hbt!]
\caption{Prompt design components for each explanation pipeline.}
\label{tab:prompt-components}
\small
\centering
\begin{tabular}{p{1.5cm} p{11.5cm}}
\toprule
Pipeline & Prompt Components \\
\midrule
Tabular- based & Predicted outcome; top-$k$ SHAP feature attributions with descriptions; instructional step-by-step reasoning; guided subgoal structure; borrower-level counterfactual prompt; output format constraint. \\[6pt]
Network- based & Predicted outcome; network connectivity analysis; instructional step-by-step reasoning for structural and relational reasoning; guided subgoal structure; network-level counterfactual prompt; output format constraint. \\[6pt]
Bimodal & Predicted outcome; top-$k$ SHAP feature descriptions; GNNExplainer subgraph descriptions; combined instructional reasoning linking borrower and relational characteristics; combined borrower-level and network-level counterfactual prompt; disambiguation hint for conflicting contributions; output format constraint. \\
\bottomrule
\end{tabular}
\end{table*}

\subsection{Fine-tuning Configurations}
\label{ft}
The final synthetic supervision dataset includes examples across all three types of pipelines (tabular-based, network-based, and bimodal). We fine-tune Gemma~3 and DeepSeek~R1 using parameter-efficient supervised instruction tuning with LoRA adapters applied to quantised base models \citep{dettmers2023qlora}. In both cases, the supervision data are formatted as system--user--assistant conversations derived from pipeline-specific evidence, and the training workflow applies chat-template formatting before tokenisation. Gemma~3 is fine-tuned using an 8-bit quantised base model with fp16 training, cosine learning-rate scheduling, and LoRA adapters applied to attention and feed-forward layers. DeepSeek~R1 uses the same overall adaptation strategy but with more conservative optimisation settings, including a smaller batch size per-device, higher gradient accumulation, shorter maximum sequence length and gradient checkpointing to accommodate its larger footprint. Figure~\ref{fig:finetune-workflow} summarises the fine-tuning workflow.

\begin{figure*}[hbt!]
\centering
\includegraphics[width=0.60\textwidth]{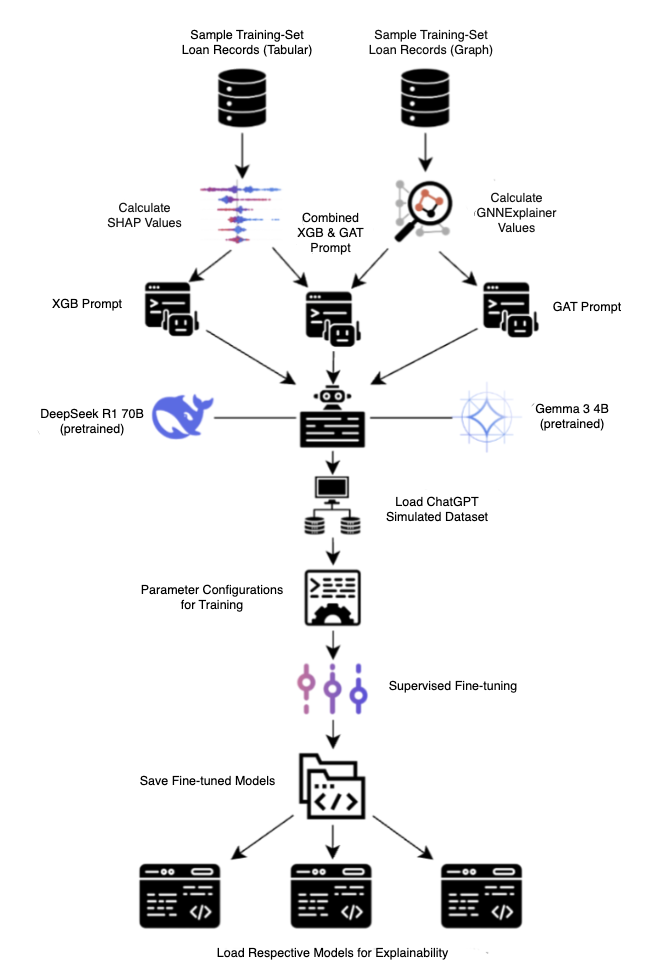}
\caption{Workflow for fine-tuning Gemma~3 and DeepSeek~R1 using LoRA-based instruction tuning on 8-bit quantised base models. SHAP and GNNExplainer outputs are paired with ChatGPT-4o-generated explanations to create the synthetic supervision dataset.}
\label{fig:finetune-workflow}
\end{figure*}

\subsection{Evaluation design and procedure}
\label{Evaluation design and procedure}
We evaluate explanation quality using a mixed design combining automated checks with human-centred assessment: automated checks identify systematic issues at scale, while human evaluation captures how explanations are interpreted in realistic decision settings. We therefore treat automated verifications as guardrails rather than substitutes for human judgement.

For automated components, we use Claude~Sonnet~4.6 \citep{anthropic2024claude} as an external judge, selected to avoid circularity with ChatGPT-4o (used for fine-tuning supervision) and Gemini~2.5 (one of the evaluated LLMs).

Human evaluation was conducted via an online survey distributed from August~27 to October~1, 2025. Each participant evaluated bimodal explanations from all three LLMs across up to three rounds (mandatory first round plus two optional), with explanation order randomised within rounds and no repeated instances per participant. Participants self-identified as Credit Risk Professionals (CRPs) or Non-Credit Risk Professionals (NCRPs) and received audience-appropriate survey materials accordingly.

\subsection{Evaluation metrics}
\label{sec:eval-metrics}
 Human evaluation captures how explanations are perceived by different stakeholder groups in realistic decision settings. For human evaluation, we operationalise the quality of the explanation along eight perceived dimensions rated on a 5-point Likert scale (1 = Strongly Disagree, 2 = Slightly Disagree, 3 = Neutral, 4 = Slightly Agree, 5 = Strongly Agree). These dimensions follow the human-centred evaluation view that explanation quality is context-dependent and should capture understanding, trust, and actionability \citep{DoshiVelez2017,miller2019explanation}. To support a risk-oriented interpretation, we treat the metrics as proxies for decision relevance in regulated credit workflows: trustworthiness and insightfulness reflect evidentiary adequacy for review and challenge; communicability reflects suitability for cross-team and customer-facing communication; and usability reflects whether the explanation could plausibly support an approval/denial workflow under operational constraints. The wording of the survey items differs slightly between CRPs and NCRPs because the two groups were presented with audience-appropriate phrasing: CRPs received domain-specific terminology aligned with credit risk workflows, whereas NCRPs received parallel plain-language wording to preserve interpretability without assuming specialist knowledge. Table~\ref{tab:eval-criteria} lists the eight dimensions and the group-specific wording used for CRPs and NCRPs.

\begin{table*}[hbt!]
\caption{Perceived explanation-quality criteria across CRPs and NCRPs.}
\label{tab:eval-criteria}
\small
\centering
\renewcommand{\arraystretch}{1.2}
\begin{tabular}{p{2.6cm} p{1.4cm} p{4.5cm} p{4.5cm}}
\toprule
Metric & Acronym & CRP Wording & NCRP Wording \\
\midrule
Understandability & UND & The explanation is easy to understand. & The explanation is easy to understand. \\
Trustworthiness & TRU & The explanation can be trusted because it presents sufficient and reliable evidence to support the loan decision. & I trust this explanation given by the system. \\
Insightfulness & INS & This explanation reveals insightful risk factors that influence loan decisions. & This explanation provides useful insight into why the decision was made. \\
Satisfaction & SAT & The level of detail satisfies my expectations for a risk assessment. & I am satisfied with how the explanation addressed my concerns about the decision. \\
Confidence & CON & This explanation increases my confidence for loan approval/denial. & I am confident in the explanation provided for the system's decision. \\
Convincingness & CVN & The justification is convincing and argues for the decision using well-weighted risk evidence. & The explanation is convincing and makes the decision seem reasonable. \\
Communicability & COM & I could use this explanation to communicate with customers and other teams in the company. & The explanation is clear enough for me to communicate with others. \\
Usability & USB & This explanation could be directly used in loan approval/denial. & I am likely to use this explanation in the future because it supports my decision-making. \\
\bottomrule
\end{tabular}
\end{table*}

\subsection{Automated proxy evaluation}
\label{Automated proxy evaluation}
We assess explanation quality using LLM-as-a-judge scores as proxies for faithfulness to model-derived evidence. Faithfulness is evaluated using Claude~Sonnet~4.6, scoring feature coverage and directional consistency against the explicit evidence blocks provided to the explanation generator \citep{Lipton2018,jacovi2020towards,zheng2023judging,gu2024survey}. Perplexity and simulated stakeholder ratings are reported as complementary proxies in \ref{appendix:D}.

\subsection{Linguistic feature analysis}
\label{Linguistic feature analysis}
To analyse which textual properties are associated with higher perceived explanation quality, we link each survey rating to the corresponding explanation text and extract a set of rule-based and library-derived linguistic features. These include length and readability measures (word count, sentence count, Flesch--Kincaid grade level, and Gunning fog index), sentiment measures (TextBlob polarity/subjectivity and VADER sentiment scores), lexical diversity, counts of numeric references, and counts of domain-relevant lexical indicators related to tabular evidence and network context. We also extract counts of hedging expressions, prescriptive language, and phrases associated with a stronger causal framing.

For each explanation-quality dimension, ratings are converted into percentile ranks and grouped into high-rated subsets (top 25\%) and low-rated subsets (bottom 25\%). We then compare the linguistic feature distributions of these groups using Cohen's \textit{d}, treating the analysis as exploratory and using effect sizes as the primary basis for interpretation.

\subsection{Reliability and statistical analysis}
\label{Reliability and statistical analysis}
Statistical methods are detailed in the \ref{appendix:D}. Briefly, human ratings use mixed-effects models with random participant intercepts; automated metrics use two-way ANOVAs with Pipeline and LLM as factors. Multiplicity is controlled via Holm correction for targeted contrasts and Benjamini--Hochberg FDR for broader pairwise comparisons.

\section{Results}
\label{Results}

This section reports results aligned with the research questions. We first provide representative examples, then report automated proxy evaluations across pipelines and LLMs, and finally present survey and linguistic analyses.

\subsection{Descriptive overview of representative explanations}
\label{Descriptive overview of representative explanations}
The three pipelines differ in attribution logic and narrative framing. Tabular-based explanations emphasise loan-level attributes with deterministic framing; network-based explanations reference cluster-level patterns but risk drifting toward unverifiable contextual claims. The bimodal pipeline integrates both evidence types into a single narrative. When tabular and relational signals align, this yields more holistic rationales; when they conflict, careful structuring is required to avoid contradictions. Representative examples of all three pipelines, including a bimodal example from Gemini~2.5, are provided in \ref{appendix:E}.

\subsection{Automated evaluation of explanation quality}
\label{Automated evaluation of explanation quality}
Addressing RQ1 and partially RQ3, this subsection reports evidence-alignment results from the LLM-as-a-judge procedure across pipelines and LLM configurations. Fluency (perplexity) and simulated audience-conditioned ratings are reported in \ref{appendix:D}, as complementary proxies; the fidelity results below represent the primary automated reference for the analysis.

\subsubsection{Evidence alignment via LLM-as-a-judge}
\label{Evidence alignment via LLM-as-a-judge}
We use an LLM-as-a-judge procedure to evaluate the evidence-grounded faithfulness of generated explanations with respect to the underlying model evidence. We operationalise faithfulness using two components: (i) feature coverage, which assesses whether the explanation references the most influential tabular or relational signals identified by the explainer, and (ii) directional consistency, which assesses whether the explanation states the correct direction of influence (e.g., risk-increasing vs.\ risk-decreasing) for those signals across both tabular and network modalities \citep{zheng2023judging, gu2024survey}. We implement this using Claude~Sonnet~4.6 as the judge LLM, scoring the following variables:

\begin{itemize}
  \item Tabular Feature Coverage (TFC): measures whether the explanation mentions the important tabular features that the predictive model actually used. Higher scores indicate that the explanation references more of the relevant features.
  \item Tabular Directional Consistency (TDC): measures whether the explanation correctly describes the direction of the effect of each tabular feature (i.e., the increase or decrease in estimated risk), as assessed by LLM-as-a-judge prompt adapted from previous work \citep{zheng2023judging, gu2024survey}.
  \item Network Feature Coverage (NFC): measures whether the explanation includes the important network-based features that influenced the model's prediction. Higher scores indicate that the explanation adheres to the correct relational evidence.
  \item Network Directional Consistency (NDC): measures whether the explanation correctly describes the direction of effect for the network features (i.e., whether a relational signal is described as positively or negatively associated with default risk), as indicated by the model’s network-based attribution signals. Higher scores therefore reflect closer alignment with the effects captured by the model.
\end{itemize}

We report all four metrics across all pipelines to verify modality specificity: pipelines that do not include a given evidence type are expected to score substantially lower on that modality’s fidelity dimensions. Table~\ref{tab:llm_judge_scores} summarises the results across pipeline--LLM configurations. The reported values are means $\pm$ 95\% confidence intervals.

\begin{table*}[hbt!]
\caption{LLM-as-a-judge scores across pipeline--LLM configurations.
Boldface indicates the highest mean per column within each pipeline.}
\label{tab:llm_judge_scores}
\small
\centering
\begin{tabular}{p{1.5cm}lcccc}
\toprule
Pipeline & LLM & TFC & TDC & NFC & NDC \\
\midrule
 \multirow{3}{1.5cm}{Tabular- based}             & Gemma~3     & 3.77\,$\pm$\,0.12 & 3.68\,$\pm$\,0.15 & 1.73\,$\pm$\,0.09 & 2.14\,$\pm$\,0.17 \\
 & DeepSeek~R1 & \textbf{4.35\,$\pm$\,0.10} & 3.48\,$\pm$\,0.13 & 1.58\,$\pm$\,0.10 & 1.91\,$\pm$\,0.17 \\
              & Gemini~2.5  & 3.91\,$\pm$\,0.11 & \textbf{3.94\,$\pm$\,0.15} & \textbf{1.90\,$\pm$\,0.09} & \textbf{2.42\,$\pm$\,0.15} \\
\midrule
 \multirow{3}{1.5cm}{Network- based}             & Gemma~3     & \textbf{1.02\,$\pm$\,0.03} & 1.10\,$\pm$\,0.07 & 4.83\,$\pm$\,0.07 & 3.20\,$\pm$\,0.09 \\
& DeepSeek~R1 & 1.00\,$\pm$\,0.00 & \textbf{1.15\,$\pm$\,0.10} & 4.46\,$\pm$\,0.11 & \textbf{3.22\,$\pm$\,0.09} \\
              & Gemini~2.5  & 1.00\,$\pm$\,0.00 & 1.01\,$\pm$\,0.02 & \textbf{4.88\,$\pm$\,0.07} & 2.89\,$\pm$\,0.10 \\
\midrule
        & Gemma~3     & 4.08\,$\pm$\,0.16 & 3.44\,$\pm$\,0.16 & 3.83\,$\pm$\,0.14 & 3.13\,$\pm$\,0.22 \\
Bimodal & DeepSeek~R1 & 4.47\,$\pm$\,0.14 & \textbf{3.67\,$\pm$\,0.18} & 3.39\,$\pm$\,0.12 & 3.36\,$\pm$\,0.11 \\
        & Gemini~2.5  & \textbf{4.73\,$\pm$\,0.13} & \textbf{3.67\,$\pm$\,0.14} & \textbf{3.98\,$\pm$\,0.16} & \textbf{3.61\,$\pm$\,0.14} \\
\bottomrule
\end{tabular}
\end{table*}

Four descriptive patterns emerge. First, tabular-based pipelines score high on tabular fidelity but near the lower bound on network metrics, consistent with the absence of relational evidence in the inputs. Within this pipeline, DeepSeek R1 achieves the highest tabular feature coverage and Gemini 2.5 the strongest tabular directional consistency.

Second, the network-based pipeline exhibits the complementary pattern. All three LLMs obtain very low scores in tabular metrics (TFC/TDC $\approx$ 1), while the coverage of the network features and the consistency of the network directionality are highest in this setting. Gemini 2.5 achieves the strongest network feature coverage, while DeepSeek R1 achieves the highest network directional consistency. These results suggest that when relational attribution signals are provided as input, the generated explanations tend to reference and correctly characterise network-based evidence.

Third, looking within each pipeline, in ten out of twelve cases feature coverage is greater than directional consistency (excluding TFC/TDC in the network-based pipeline and NFC/NDC in the tabular-based pipeline since the pipeline can not have the respective evidence in these cases). 
In the network-based pipeline, NFC is in the range $4.46-4.88$ whereas NDC is in the range $2.89-3.22$ meaning there is a gap of $1.24-1.99$ between the two measures. Similarly, in the bimodal pipeline, TDC stalls at 3.67, whereas TFC reaches 4.73. This indicates that the narratives reliably name the right risk factors, but are less reliable when stating whether each one pushes risk up or down. 

Finally, the bimodal pipeline produces comparatively balanced scores in both modalities. In the bimodal pipeline, Gemini 2.5 achieves the highest tabular feature coverage and network feature coverage, while DeepSeek R1 and Gemini 2.5 achieve equal tabular directional consistency (3.67 ± 0.18 and 3.67 ± 0.14, respectively). This pattern suggests that when both tabular and relational attribution signals are available to the explanation generator, the resulting narratives can reference both evidence types simultaneously, reducing the modality-specific trade-offs observed in single-modality pipelines. To quantify systematic differences, we performed two-way ANOVAs for each fidelity metric with the factors Pipeline and LLM, including their interaction (Table~\ref{tab:anova_judge}).

\begin{table}[hbt!]
\caption{Two-way ANOVA results for LLM-as-a-judge fidelity metrics.
All effects are statistically significant ($p < 10^{-6}$).}
\label{tab:anova_judge}
\small
\centering
\begin{tabular}{lcccccc}
\toprule
Metric
& $F_{\text{pipe}}$ &  $\eta^2_{\text{pipe}}$
& $F_{\text{llm}}$  & $\eta^2_{\text{llm}}$
& $F_{\text{int}}$ & $\eta^2_{\text{int}}$ \\
\midrule
TFC & 328.96  & 0.391
    & 11.81 & 0.014
    & 10.04  & 0.024 \\
TDC & 299.02  & 0.352
    & 26.08  & 0.031
    & 22.07  & 0.052 \\
NFC & 354.27  & 0.377
    & 62.95  & 0.067
    & 21.66  & 0.046 \\
NDC & 39.19  & 0.069
    & 28.94  & 0.051
    & 11.63  & 0.041 \\
\bottomrule
\end{tabular}
\end{table}

The pipeline is the dominant source of variance for tabular feature coverage, tabular directional consistency, and network coverage, accounting for approximately 35–39\% of the total variance across these metrics. For network directional consistency (whether explanations correctly capture the direction of network-based attribution signals), the variance is more evenly distributed between the pipeline, LLM, and their interaction, indicating that both evidence modality and the choice of LLM contribute to the differences in the directional alignment.

The choice of LLM explains a smaller but statistically significant share of variance across all four metrics, particularly for network coverage ($\eta^2=0.067$) and tabular directional consistency ($\eta^2=0.031$). The Pipeline$\times$LLM interaction terms are also significant for all metrics, indicating that the differences between LLMs depend on the type of evidence available to the explanation generator.

\subsection{Human evaluation of explanation quality}
\label{Human evaluation of explanation quality}
This subsection examines differences in perceived explanation quality between CRPs ($n=48$) and NCRPs ($n=101$), evaluating bimodal explanations across the eight dimensions in Subsection~\ref{sec:eval-metrics}. The human study was restricted to the bimodal pipeline to reduce participant burden while focusing on the most demanding evaluation condition, since it combines both tabular and network evidence. Consequently, conclusions about LLM-level differences should be interpreted within the bimodal pipeline and may not generalise uniformly to single-modality pipelines. The characteristics of the participants are summarised in \ref{app:demographics}.

\subsubsection{Ratings collected via survey}
\label{Ratings collected via survey}
Tables~\ref{tab:crp_bimodal_matrix} and~\ref{tab:ncrp_bimodal_matrix} summarise item-level mean Likert ratings for bimodal explanations by cohort and LLM in the eight dimensions in Subsection~\ref{sec:eval-metrics}. The reported values are means $\pm$ 95\% confidence intervals, computed as $\bar{x} \pm t_{0.975,\,n-1}\cdot \mathrm{SEM}$ with $\mathrm{SEM}=\mathrm{SD}/\sqrt{n}$.
\begin{table*}[hbt!]
\caption{Human (CRP) mean Likert ratings across explanation-quality
metrics. Boldface indicates the highest mean per column.}
\label{tab:crp_bimodal_matrix}
\small
\centering
\begin{tabular}{lccc}
\toprule
Metric & Gemma~3 & DeepSeek~R1 & Gemini~2.5 \\
\midrule
UND & \textbf{4.65\,$\pm$\,0.14} & 4.24\,$\pm$\,0.23 & 4.41\,$\pm$\,0.19 \\
TRU & \textbf{4.23\,$\pm$\,0.21} & 3.93\,$\pm$\,0.25 & 3.98\,$\pm$\,0.22 \\
INS & \textbf{4.09\,$\pm$\,0.25} & 3.78\,$\pm$\,0.25 & 3.88\,$\pm$\,0.23 \\
SAT & \textbf{3.97\,$\pm$\,0.24} & 3.75\,$\pm$\,0.29 & 3.57\,$\pm$\,0.27 \\
CON & \textbf{4.10\,$\pm$\,0.22} & 3.70\,$\pm$\,0.27 & 3.87\,$\pm$\,0.24 \\
CVN & \textbf{4.08\,$\pm$\,0.22} & 3.64\,$\pm$\,0.29 & 3.74\,$\pm$\,0.25 \\
COM & \textbf{4.42\,$\pm$\,0.19} & 3.64\,$\pm$\,0.29 & 3.98\,$\pm$\,0.24 \\
USB & \textbf{3.99\,$\pm$\,0.20} & 3.39\,$\pm$\,0.29 & 3.40\,$\pm$\,0.29 \\
\bottomrule
\end{tabular}
\end{table*}

\begin{table*}[hbt!]
\caption{Human (NCRP) mean Likert ratings across explanation-quality
metrics. Boldface indicates the highest mean per column.}
\label{tab:ncrp_bimodal_matrix}
\small
\centering
\begin{tabular}{lccc}
\toprule
Metric & Gemma~3 & DeepSeek~R1 & Gemini~2.5 \\
\midrule
UND & \textbf{4.36\,$\pm$\,0.11} & 4.22\,$\pm$\,0.15 & 4.18\,$\pm$\,0.17 \\
TRU & \textbf{4.24\,$\pm$\,0.11} & 4.19\,$\pm$\,0.14 & 4.01\,$\pm$\,0.16 \\
INS & \textbf{4.11\,$\pm$\,0.13} & 4.03\,$\pm$\,0.15 & 4.06\,$\pm$\,0.16 \\
SAT & 4.08\,$\pm$\,0.14 & \textbf{4.12\,$\pm$\,0.14} & 4.01\,$\pm$\,0.17 \\
CON & \textbf{4.24\,$\pm$\,0.12} & 4.17\,$\pm$\,0.15 & 4.11\,$\pm$\,0.16 \\
CVN & \textbf{4.14\,$\pm$\,0.13} & 4.01\,$\pm$\,0.15 & 4.04\,$\pm$\,0.16 \\
COM & \textbf{4.20\,$\pm$\,0.11} & 3.97\,$\pm$\,0.16 & 3.92\,$\pm$\,0.15 \\
USB & \textbf{4.20\,$\pm$\,0.14} & 3.82\,$\pm$\,0.19 & 4.14\,$\pm$\,0.16 \\
\bottomrule
\end{tabular}
\end{table*}

To address RQ2, we compare mean ratings and confidence intervals between CRPs and NCRPs. Descriptively, Gemma~3 received the highest mean ratings across all eight metrics for CRPs and seven of eight for NCRPs, with the exception of satisfaction (SAT), where DeepSeek~R1 scored marginally higher.

With respect to RQ3, these patterns suggest that the smaller fine-tuned Gemma~3 was often perceived more favourably than the larger zero-shot Gemini~2.5 in dimensions such as understandability, trustworthiness, and usability. However, because confidence intervals overlap between LLMs, we interpret these differences as descriptive tendencies rather than statistically reliable gaps.

The cohort patterns also differ. CRPs show clearer separation across LLMs, whereas NCRPs rate explanations more uniformly, suggesting that domain expertise shapes how users distinguish between explanation styles. In particular, CRPs appear to apply stricter evidentiary standards, which leads to greater differentiation between LLMs. Figures~\ref{fig:gemma3-cohort}--\ref{fig:gemini-cohort} illustrate these between-cohort patterns across the eight evaluation metrics.

\begin{figure}[hbt!]
\centering
\subfloat[Gemma~3\label{fig:gemma3-cohort}]{%
  \includegraphics[width=0.45\linewidth]{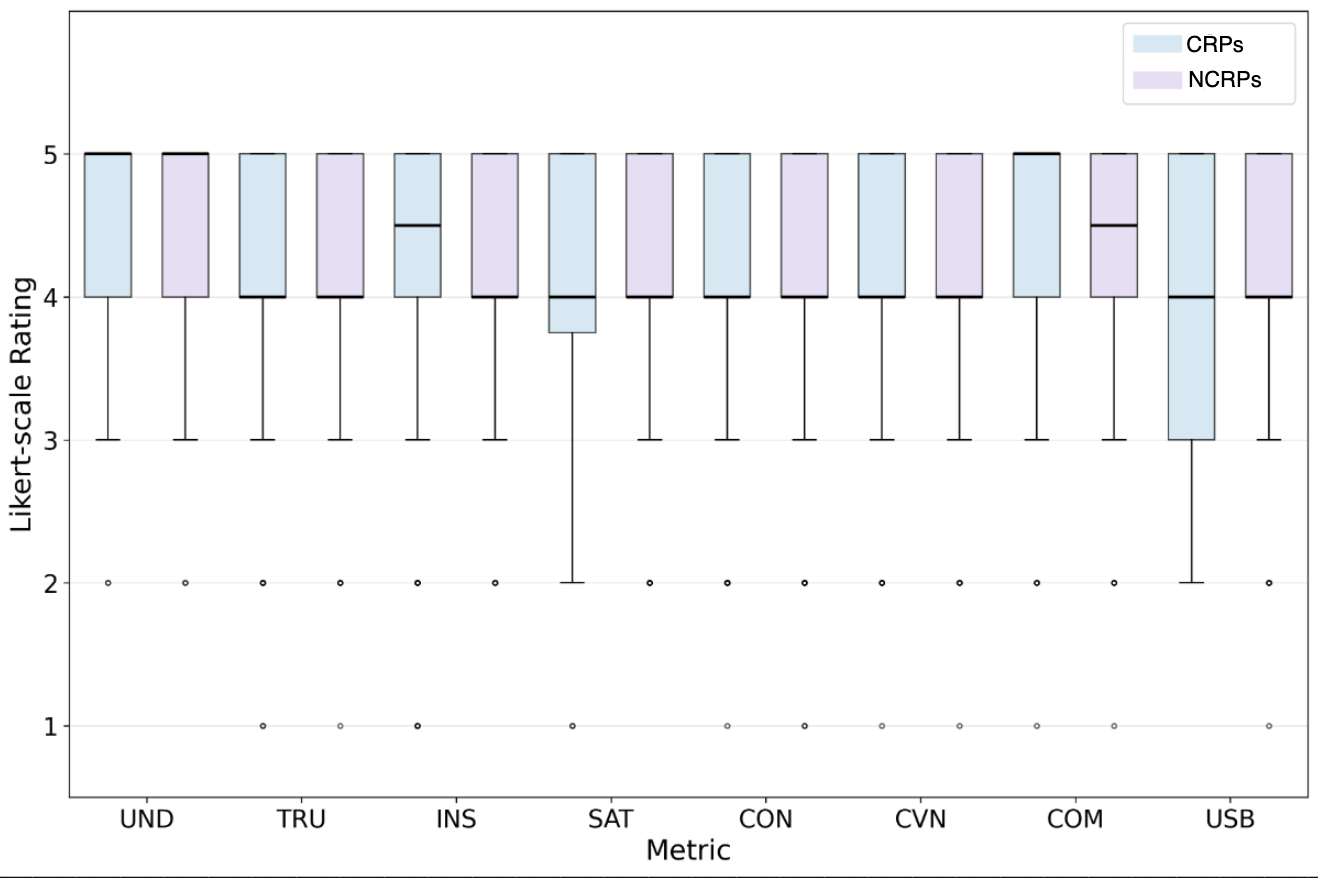}} 
\subfloat[DeepSeek~R1\label{fig:deepseek-cohort}]{%
  \includegraphics[width=0.45\linewidth]{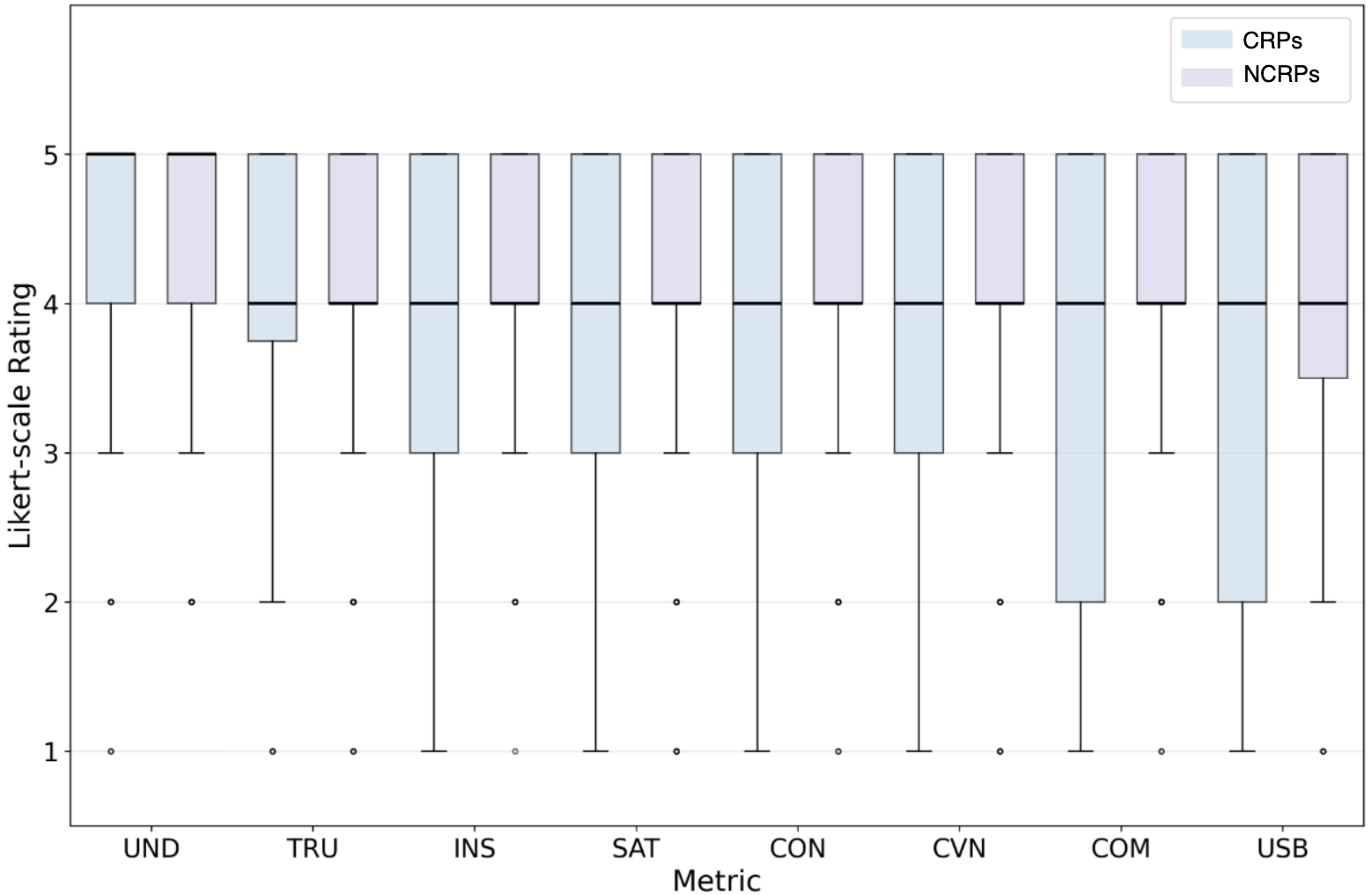}}\\[2pt]
\subfloat[Gemini~2.5\label{fig:gemini-cohort}]{%
  \includegraphics[width=0.45\linewidth]{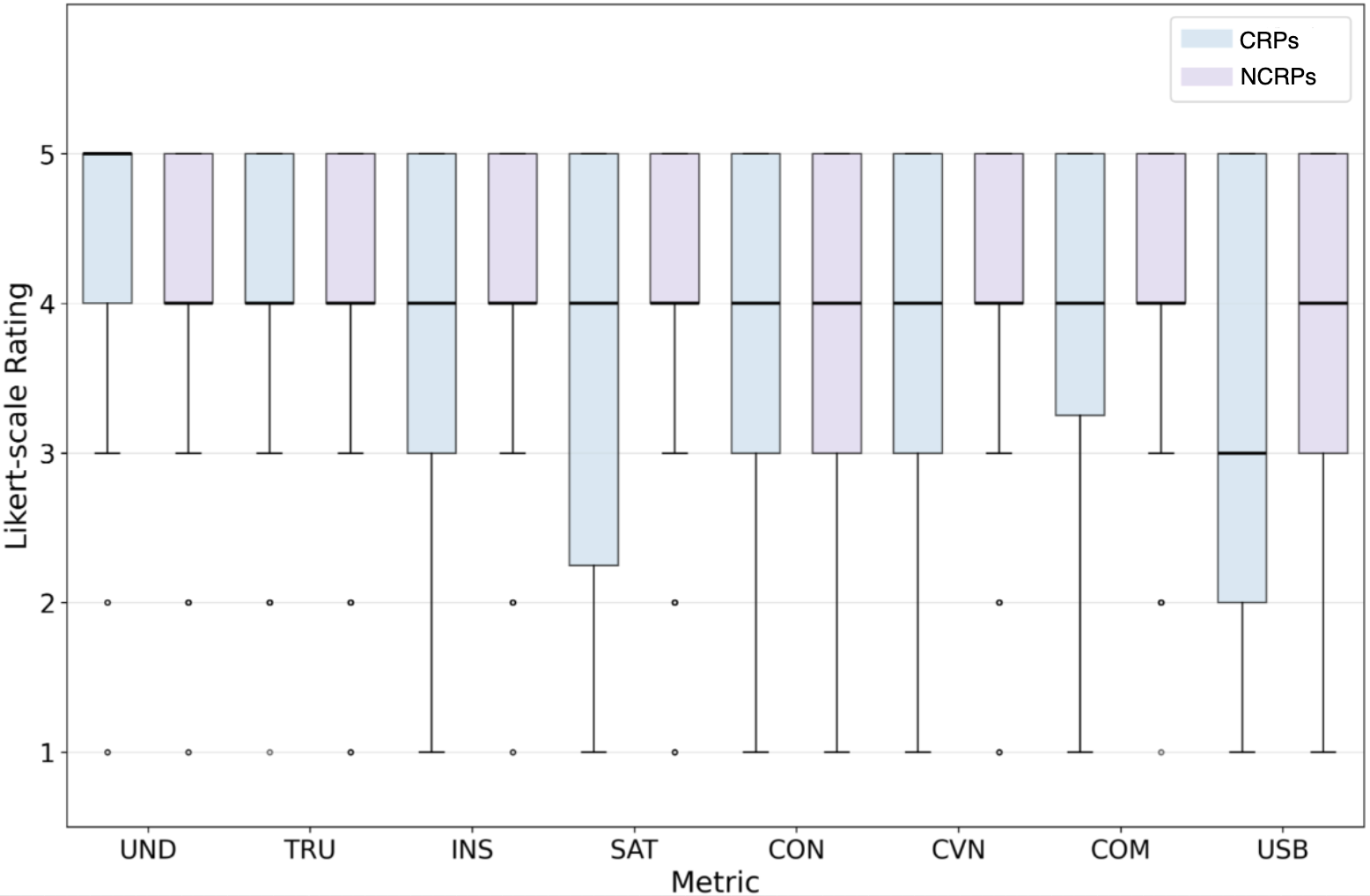}}
\caption{Between-cohort comparisons by LLM.}
\label{fig:ft-cohort}
\end{figure}

Across LLMs and metrics, cohort-level differences are modest and not consistently directional. NCRPs sometimes report slightly higher mean ratings, but the pattern is not uniform across LLMs or dimensions, and median ratings often coincide at 4 or 5. We therefore rely primarily on mean ratings with confidence intervals for descriptive comparison. In general, both CRPs and NCRPs positively rate the bimodal explanations, suggesting acceptable interpretability for both cohorts in this setting.

Several LLM-level patterns also emerge. Gemma~3 receives the highest CRP rating in UND and COM, suggesting that CRPs found its explanations easier to follow and more suitable for communication within credit workflows. USB shows the largest CRP--NCRP gap, with CRPs rating all LLMs lower; this may reflect greater caution about operational use of narrative explanations. NCRPs also report slightly higher SAT scores in LLMs, including DeepSeek~R1. These descriptive patterns provide context for the mixed-effects analyses reported in Subsection~\ref{Mixed-effects analysis of survey ratings}.

\subsubsection{Mixed-effects analysis of survey ratings}
\label{Mixed-effects analysis of survey ratings}
This subsection reports the results for the human survey, comparing perceived explanation quality across the three LLMs, two cohorts (CRP vs.\ NCRP), and eight decision-relevant metrics. Because participants could opt in to additional rounds, the number of ratings per participant and per LLM is not constant. We therefore summarise participant-level mean Likert ratings (1--5; higher is better) and standard deviations, and use mixed-effects modelling to account for repeated measures and rater-specific leniency or strictness. Table~\ref{tab:means-by-cohort} reports participant-level mean ratings by cohort, LLM, and metric.

\begin{table*}[hbt!]
\small
\centering
\setlength{\tabcolsep}{4pt}
\renewcommand{\arraystretch}{1.15}
\caption{Participant-level mean Likert ratings by cohort, LLM, and metric, with standard deviations in parentheses}
\label{tab:means-by-cohort}
\begin{tabular}{lcccccc}
\toprule
& \multicolumn{3}{c}{CRPs} 
& \multicolumn{3}{c}{NCRPs} \\
\cmidrule(lr){2-4} \cmidrule(lr){5-7}
Metric 
& \makecell{Gemma~3} 
& \makecell{DeepSeek~R1} 
& \makecell{Gemini~2.5}
& \makecell{Gemma~3} 
& \makecell{DeepSeek~R1} 
& \makecell{Gemini~2.5} \\
\midrule
UND & 4.05 (0.87) & 3.90 (0.88) & 3.94 (0.88) & 4.10 (0.76) & 3.78 (1.04) & 4.07 (0.85) \\
TRU   & 4.17 (0.82) & 3.92 (1.00) & 3.68 (1.05) & 4.30 (0.60) & 4.01 (0.73) & 4.07 (0.78) \\
INS   & 4.31 (0.82) & 3.77 (1.08) & 3.68 (0.99) & 4.39 (0.55) & 4.14 (0.67) & 4.07 (0.88) \\
SAT      & 4.40 (0.67) & 3.81 (1.03) & 3.65 (0.93) & 4.38 (0.62) & 4.16 (0.71) & 4.09 (0.79) \\
CON       & 4.35 (0.71) & 3.62 (1.16) & 3.68 (0.92) & 4.36 (0.58) & 4.11 (0.72) & 4.15 (0.83) \\
CVN   & 4.45 (0.71) & 3.85 (1.03) & 3.77 (0.87) & 4.34 (0.65) & 4.12 (0.75) & 4.13 (0.79) \\
COM   & 4.19 (0.92) & 3.66 (1.09) & 3.80 (0.89) & 4.14 (0.80) & 3.94 (0.93) & 3.79 (1.06) \\
USB        & 3.91 (0.94) & 3.22 (1.22) & 3.47 (1.12) & 3.75 (0.97) & 3.67 (0.91) & 3.71 (0.95) \\
\bottomrule
\end{tabular}
\end{table*}

Two descriptive patterns emerge. First, Gemma~3 has the highest mean ratings across all eight metrics in both cohorts, although the overlapping uncertainty suggests that these should be interpreted as tendencies rather than reliable performance differences. This pattern is consistent with, but does not establish, the possibility that domain-aligned fine-tuning improves perceived explanation quality relative to a larger zero-shot commercial LLM. The overlapping uncertainty intervals reflect the modest sample sizes and considerable within-group variability in Likert ratings, which together limit statistical power. Second, the differences between DeepSeek~R1 and Gemini~2.5 are generally small, whereas the differences between the cohorts are more apparent in operationally grounded dimensions such as trustworthiness and usability, where the CRPs tend to rate both LLMs lower than the NCRPs. This is consistent with prior work showing that the quality of the explanation depends on the context of the stakeholders and that NCRPs may respond differently to the differences in presentation \citep{DoshiVelez2017,miller2019explanation,peters2008numeracy,dieckmann2009use}.

\paragraph{Mixed-effects model and cohort sensitivity}
To formally test these patterns, we fit a mixed-effects model (formula: \texttt{score \textasciitilde{} cohort * llm * metric + (1 | participant\_id)}) to participant-level mean ratings, treating Cohort as a between-subject factor, LLM and Metric as within-subject factors, and including a random intercept for participant to capture stable rater leniency/strictness. Three nested likelihood-ratio tests (fitted by ML) assess the contribution of cohort-related terms. Adding all cohort terms significantly improves fit relative to a model with only LLM and Metric effects ($\chi^2 = 88.20$, $df = 24$, $p < 0.001$), indicating systematic differences between CRPs and NCRPs. Adding cohort-by-LLM terms (and associated higher-order terms) yields a further improvement ($\chi^2 = 36.40$, $df = 16$, $p = 0.003$), consistent with the descriptive pattern that CRPs penalise the non-Gemma~3 LLMs more strongly than NCRPs. For robustness, we tested the three-way interaction LLM~$\times$~Cohort~$\times$~Metric using a Type~III Wald test on the full model; the result is not significant ($F = 1.34$, $df = 14$, $p = 0.178$), suggesting that cohort differences in LLM ratings are broadly consistent across metrics rather than driven by a single dimension.

\paragraph{Multiplicity control and pairwise contrasts}
Given eight metrics and multiple model comparisons, we report effect sizes and uncertainty intervals as primary summaries and use $p$-values as complementary indicators, applying Benjamini--Hochberg false-discovery-rate control for confirmatory contrasts. Because participants evaluated up to two of the three models per round, paired contrasts between DeepSeek~R1 and Gemini~2.5 are based on the subset who rated both models within each cohort (CRPs: $n = 10$; NCRPs: $n = 11$). To make DeepSeek~R1 vs.\ Gemini~2.5 differences concrete, Table~\ref{tab:deepseek-vs-gemini} reports Holm-corrected and Benjamini--Hochberg FDR corrected paired contrasts within each cohort (positive values indicate higher ratings for DeepSeek~R1). Under Holm correction, no contrast reaches significance (minimum $p_{\mathrm{Holm}} = 0.053$, NCRPs COM); under Benjamini--Hochberg FDR control the result is identical (minimum $p_{\mathrm{BH}} = 0.053$). The NCRPs COM contrast represents a descriptive tendency toward higher communicability ratings for Gemini~2.5 among NCRPs, but does not meet the threshold for a reliable separation. We therefore treat all DeepSeek~R1 vs.\ Gemini~2.5 differences as small and context-dependent rather than reliable separations.

\begin{table*}[hbt!]
\small
\centering
\setlength{\tabcolsep}{5pt}
\renewcommand{\arraystretch}{1.15}
\caption{Paired differences between DeepSeek~R1 and Gemini~2.5 by metric 
and cohort, with Holm-adjusted and Benjamini--Hochberg FDR-adjusted 
$p$-values. Positive values indicate higher ratings for DeepSeek~R1.}
\label{tab:deepseek-vs-gemini}
\begin{tabular}{lrrrrrrrrr}
\toprule
& \multicolumn{4}{c}{CRPs} 
& \multicolumn{4}{c}{NCRPs} \\
\cmidrule(lr){2-5} \cmidrule(lr){6-9}
Metric 
& Mean diff. & SD diff. & $p_{\mathrm{Holm}}$ & $p_{\mathrm{BH}}$
& Mean diff. & SD diff. & $p_{\mathrm{Holm}}$ & $p_{\mathrm{BH}}$ \\
\midrule
UND & $-$0.050 & 0.468 & 1.000 & 0.681 
    & $-$0.121 & 0.583 & 1.000 & 0.733 \\
TRU & $+$0.350 & 0.489 & 0.250 & 0.250 
    & $+$0.075 & 0.657 & 1.000 & 0.765 \\
INS & $+$0.150 & 0.648 & 1.000 & 0.732 
    & $-$0.045 & 0.510 & 1.000 & 0.765 \\
SAT & $+$0.250 & 0.635 & 1.000 & 0.517 
    & $-$0.288 & 0.455 & 0.722 & 0.413 \\
CON & $+$0.250 & 0.687 & 1.000 & 0.517 
    & $-$0.150 & 0.555 & 1.000 & 0.733 \\
CVN & $+$0.300 & 0.665 & 1.000 & 0.517 
    & $-$0.068 & 0.560 & 1.000 & 0.765 \\
COM & $+$0.200 & 0.806 & 1.000 & 0.517 
    & $-$0.500 & 0.512 & 0.053 & 0.053 \\
USB & $+$0.275 & 0.845 & 1.000 & 0.532 
    & $-$0.152 & 0.351 & 1.000 & 0.725 \\
\bottomrule
\end{tabular}
\end{table*}

\paragraph{Variance decomposition}
Finally, Table~\ref{tab:variance-decomp} decomposes the variance in overall explanation ratings into participant- and explanation-case components. Participant-level strictness or leniency accounts for 46.3\% of the variance, supporting the inclusion of random participant intercepts. Explanation cases account for 10.0\%, indicating a meaningful contribution from the specific loans or scenarios being explained, even under fixed prompting rules. The remaining 43.7\% is residual variance, reflecting within-participant variability not captured by stable rater tendencies or case-level effects.

\begin{table}[!hbt]
\small
\centering
\caption{Variance decomposition of overall explanation ratings from a 
random-intercepts model with participant and explanation-instance effects.}
\label{tab:variance-decomp}
\begin{tabular}{lrr}
\toprule
Component & Variance & Share (\%) \\
\midrule
Participants (leniency/strictness) & 0.507 & 46.3 \\
Explanation cases (LLM$\times$instance) & 0.110 & 10.0 \\
Residual & 0.478 & 43.7 \\
\bottomrule
\end{tabular}
\end{table}

Overall ratings are positive, cohort differences are systematic but modest, and Gemma 3 shows a descriptive advantage; overlapping intervals caution against treating these as definitive separations.

\subsubsection{Linguistic and sentiment feature analysis}
\label{Linguistic and sentiment feature analysis}
This subsection addresses RQ4 by linking textual properties with human-rated explanation quality in the bimodal pipeline. We compare linguistic feature profiles of higher- versus lower-rated explanations to identify which characteristics are associated with perceived quality. For each explanation-quality dimension, we define high-rated explanations as the top 25\% and low-rated explanations as the bottom 25\% based on human ratings, to maximise contrast between groups. We then extract 16 predefined linguistic features that span emotional tone, structural complexity, and domain-specific financial language (Table~\ref{tab:linguistic-features}).

\begin{table*}[hbt!]
\caption{Linguistic feature categories used in the textual analysis.}
\label{tab:linguistic-features}
\small
\centering
\begin{tabular}{p{1.4cm} l p{7cm}}
\toprule
Category & Feature & Description \\
\midrule
\multirow{6}{*}{Sentiment}
& sentiment\_polarity & TextBlob polarity score $\in [-1, 1]$ (negative to positive) \\
& sentiment\_subjectivity & TextBlob subjectivity $\in [0, 1]$ (objective to subjective) \\
& vader\_compound & VADER composite sentiment score $\in [-1, 1]$ \\
& vader\_pos & VADER positive sentiment proportion \\
& vader\_neg & VADER negative sentiment proportion \\
& vader\_neu & VADER neutral sentiment proportion \\
\midrule
\multirow{6}{*}{Structural}
& word\_count & Total number of words in explanation \\
& sentence\_count & Total number of sentences in explanation \\
& flesch\_kincaid & Flesch--Kincaid readability grade level (higher = more complex) \\
& gunning\_fog & Gunning Fog readability index (higher = more complex) \\
& unique\_tokens & Number of distinct words (lexical diversity) \\
& shap\_feature\_count & Mentions of SHAP feature names per 100 words \\
\midrule
\multirow{4}{1.2cm}{Domain- specific}
& numeric\_count & Frequency of numeric values per 100 words \\
& network\_term\_count & Credit/network terminology per 100 words \\
& hedge\_count & Hedging expressions (``may'', ``possibly'', ``suggests'') per 100 words \\
& causal\_overclaiming\_count & Strong causal language (``proves'', ``causes'', ``demonstrates that'') per 100 words \\
\bottomrule
\end{tabular}
\end{table*}

We quantify differences in feature prevalence between high- and low-rated groups using Cohen's $d$ effect sizes, computed separately for each quality dimension (Figure~\ref{fig:linguistic-heatmap}). Positive values indicate features that are more prevalent in high-rated explanations; negative values indicate features associated with lower ratings.

\begin{figure*}[hbt!]
\centering
\includegraphics[width=0.75\textwidth]{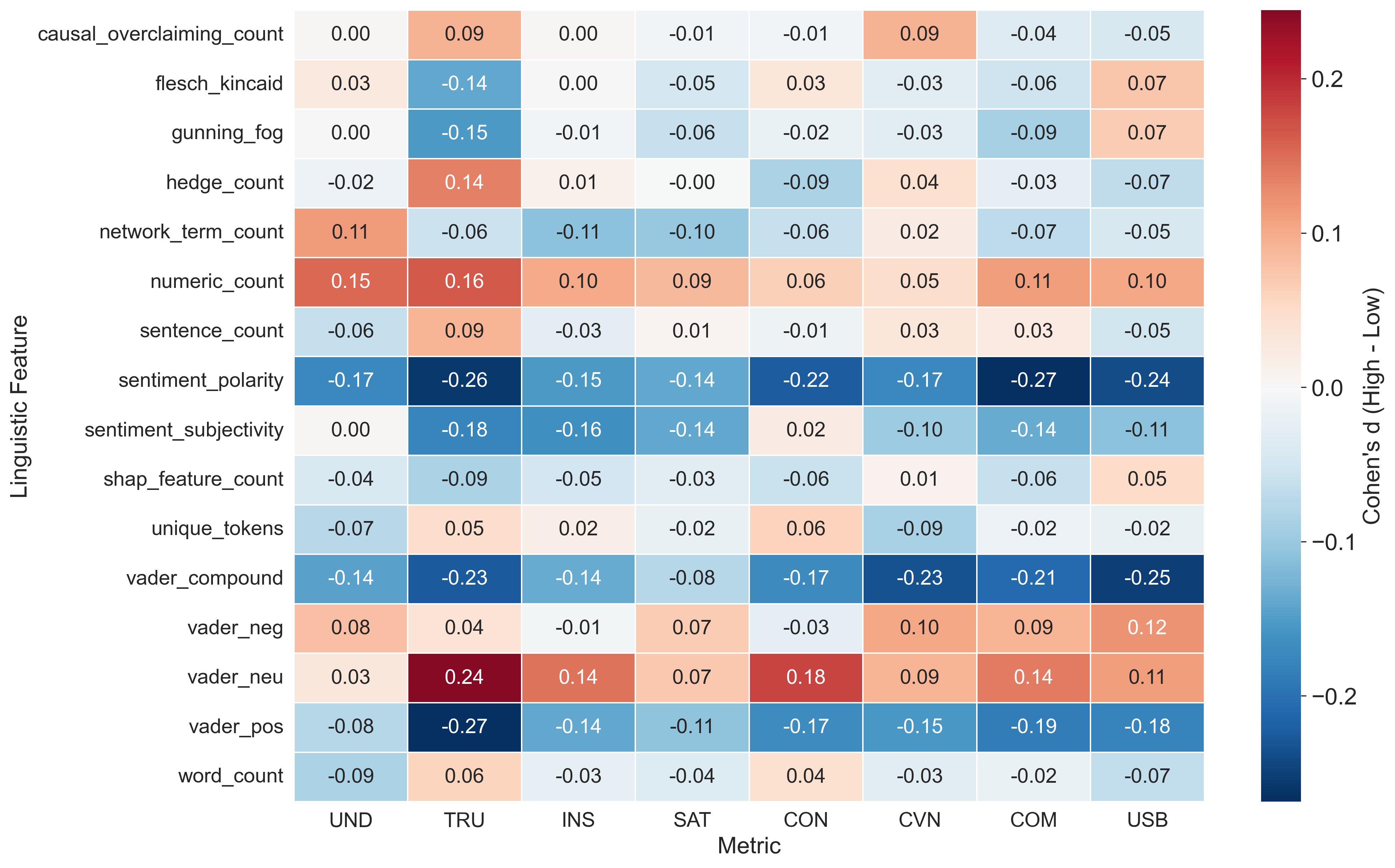}
\caption{Cohen's $d$ effect sizes comparing linguistic features in high-rated (top quartile) and low-rated (bottom quartile) explanations across quality metrics. Positive values (red) indicate features more prevalent in high-rated explanations; negative values (blue) indicate the reverse.}
\label{fig:linguistic-heatmap}
\end{figure*}

Three patterns emerge (Figure~\ref{fig:linguistic-heatmap}). First, higher-rated explanations use a more neutral, factual tone: trustworthiness is associated with higher neutral sentiment (\texttt{vader\_neu}, $d=0.24$) and lower positive emotional language (\texttt{sentiment\_polarity}, $d=-0.26$). Similar effects hold for communicability and usability. Second, numeric density (\texttt{numeric\_count}, $d\approx0.05$--$0.16$) is positively associated with perceived quality across dimensions, suggesting evaluators prefer inspectable values over qualitative descriptions. Third, stylistic complexity (readability indices, lexical diversity, hedging) shows negligible associations ($|d|\leq0.15$).

Together, these results provide practical guidance: high-quality credit risk explanations should prioritise neutral tone and concrete numeric evidence while avoiding subjective or emotionally framed language.

\section{Discussion}
\label{Discussion}

This section interprets the findings in relation to the RQs, focusing on their implications for credit risk communication, the underlying mechanisms, and the practical and policy limitations.

\subsection{Impact on credit risk assessment and communication}
\label{Impact on credit risk assessment and communication}
With respect to RQ1, explanation modality is the dominant determinant of evidence-grounded fidelity in our automated evaluations. Evidence modality, operationalised through pipeline structure, explains substantially more variance in feature coverage and directional consistency than LLM choice, which nonetheless plays a secondary but statistically significant role, most pronounced for network directional consistency (Subsection~\ref{Evidence alignment via LLM-as-a-judge}).

In terms of RQ3, the apparent benefits of fine-tuning are pipeline-dependent. Fine-tuned LLMs show the clearest advantages in the tabular-based pipeline, whereas the zero-shot LLM remains competitive in bimodal and network-based pipelines. In the human study, domain alignment is associated with stronger perceived operational adequacy, particularly among CRP evaluators whose criteria align more closely with audit and validation needs. NCRPs also tend to rate Gemma~3 higher in communicability and usability, suggesting that domain alignment can benefit both cohorts in some settings (Subsections~\ref{Evidence alignment via LLM-as-a-judge} and~\ref{Human evaluation of explanation quality}). At the same time, descriptive examples show that network-based narratives can drift toward plausible but unverifiable contextual statements (see \ref{appendix:E}), which motivates tighter grounding constraints in regulated settings.

The results in Subsection~\ref{Evidence alignment via LLM-as-a-judge} furthermore reveal a systematic asymmetry between what the explanations cover and their directional consistency, which is considerably less reliable than feature coverage. 
In a regulated credit context this presents a severe risk, as an adverse-action communication that names a correct variable with an inverted sign is a compliance
breach, not a stylistic flaw, while being invisible to a reader who cannot access the attribution. 
A fluency-oriented evaluation would score such an explanation highly but it is the directional error, not omission, that is the governance-relevant failure, further highlighting the need for evaluation standards for stakeholder-appropriate explanations \citep{de2024explainable}.

The relatively strong performance of the smaller domain-adapted LLM further suggests that perceived explanation quality depends less on parameter count than on alignment between the LLM, the evidence representation, and the target use case, which is consistent with regulatory guidance emphasising that decisioning systems should be transparent, well understood, and fit for purpose \citep{OSFI2024ModelRisk}.
That a 4-billion-parameter model matched or exceeded a 70-billion-parameter model and a commercial API on perceived explanation quality carries an efficiency implication: for this task, adequate explanation quality does not require frontier-scale models. Smaller domain-aligned models are substantially cheaper to run per inference and correspondingly lower in energy footprint, and can be hosted on-premise rather than called through an external API. In a setting where borrower-facing explanations may be generated at high volume, this couples the governance advantages of domain alignment with a materially lower computational and environmental cost.

\subsection{Mechanisms underlying the observed effects}
\label{Mechanisms underlying the observed effects}
Two complementary mechanisms help explain why domain-aligned LLMs performed more strongly and why NCRPs tended to rate explanations more positively. First, fine-tuned LLMs were trained in supervision aligned with the target domain and the evidence-to-text format used in our pipelines. This encourages narratives to prioritise inspectable evidentiary cues (e.g., debt-to-income ratios, credit scores, and network statistics) over generic persuasive framing. Consistent with RQ4, the linguistic analysis in Subsection~\ref{Linguistic and sentiment feature analysis} shows that higher-rated explanations are more neutral and evidence-forward: trustworthiness is associated with higher neutral sentiment and lower positively valenced language, while numeric specificity is positively associated with perceived quality. Together, these patterns suggest that domain alignment promotes factual, evidence-grounded statements rather than persuasive narratives.

Second, and central to RQ2, CRP and NCRP evaluators appear to apply different heuristics when judging explanation quality (see Subsections~\ref{Human evaluation of explanation quality} and~\ref{sec:eval-metrics}). Research on the affect heuristic suggests that people often rely on emotional responses when evaluating risk, with positive affect reducing perceived risk and negative affect increasing caution \citep{Slovic2007Affect}. Related work in behavioural decision-making shows that vivid and emotionally salient information can exert a disproportionate influence on risk judgments compared to purely statistical information \citep{Loewenstein2001Risk}. In our study, NCRPs tended to rate explanations more highly when they included narrative context and conversational framing, even when evidentiary linkage was weaker, whereas CRPs were more likely to down-rate explanations lacking precise, inspectable numerical justification. This divergence is consistent with a governance-oriented view of explainability: affective framing can increase perceived communicability while weakening trustworthiness and operational adequacy, a pattern that general-purpose LLMs favouring fluent evaluative language may amplify.

\subsection{Implications and limitations for practice and policy}
\label{Implications and limitations for practice and policy}
These mechanisms have practical implications for governance. For example, the Canadian Banking Regulator's model risk guidance emphasises transparency, documentation, independent challenge, and ongoing monitoring within a risk-based framework \citep{OSFI2024ModelRisk}. Consistent with these priorities, on-premise fine-tuned LLMs may offer greater auditability and change control than commercial API-based systems, although we do not test governance properties directly. In contrast, zero-shot commercial LLMs often provide limited visibility into training data, update schedules, and drift, increasing compliance exposure.

Second, organisations should exercise caution when using LLMs for adverse-action letters: creditors remain responsible for specific evidence-based reasons and cannot rely on generative systems to obscure or generalise them \citep{cfpb2022circular}, and our findings suggest generic LLMs may produce fluent but weakly grounded narratives that omit key quantitative evidence. Borrower-facing explanations should therefore rely on auditable, domain-aligned systems with logs linking each narrative to its supporting SHAP/GNN artefacts and explainer/LLM versions; LLMs with opaque training and update processes should not be deployed without extensive internal validation and documentation. Fairness is an important direction for future work: standard generative LLMs may behave differently across demographic groups \citep{Barocas2023Fairness, kozodoi2022fairness}, so future studies should evaluate explanation quality across protected borrower characteristics.

These implications should be interpreted in light of several limitations. First, our tasks focus on consumer credit in a single jurisdiction, which may limit generalisability, although this is partially mitigated by a survey that included participants from multiple jurisdictions. Second, outputs are sensitive to prompt phrasing and stochastic decoding, and provider updates can induce behavioural drift, underscoring the need for continuous governance oversight. Third, residual dependence from shared prompts and training distributions cannot be fully excluded and may affect variance estimates. Finally, evaluator heterogeneity introduces noise; the CRP vs.\ NCRP differences are informative but may not represent the full stakeholder set.

\section{Conclusions}
\label{Conclusions}

This study examined how LLMs can function as explanation layers in credit risk modelling, particularly whether they can translate technical explanation artefacts into narratives that are faithful, decision-relevant and usable in practice. We observe systematic differences between CRP and NCRP evaluators: CRPs apply stricter evidentiary and operational criteria, whereas NCRPs place greater weight on narrative clarity and communicability. The LLM configuration also matters. Across the evaluated configurations, domain-adapted LLMs often align strongly with structured tabular evidence, especially in tabular-based pipelines, although the zero-shot LLM remains competitive in several bimodal metrics. Across configurations, the explanation modality is the dominant driver of automated fidelity, indicating that the evidence provided to the explainer fundamentally shapes what can be communicated and audited. We also identify a systematic asymmetry between what generated explanations cover and what they get directionally right, which can have great consequences for regulated credit communications.

Our linguistic analysis suggests that higher-rated explanations tend to use a more neutral tone and include more concrete numeric detail, although these associations are modest. Taken together, the findings indicate that both the LLM choice and the evidence modality shape the quality of the explanation and therefore warrant explicit governance attention. In regulated credit settings, institutions should prioritise domain-aligned explanation systems, maintain robust evaluation and monitoring, and retain evidence logs linking narratives to the underlying explanation artefacts. More broadly, combining regulatory guidance, human judgement, and tailored machine learning offers a practical path toward risk-aware AI in financial decision-making.

Future research could extend this framework across jurisdictions, lending products, and languages to assess external validity, and expand evaluation cohorts to include regulators, underwriters, and borrowers to clarify stakeholder-specific requirements. Developing benchmark datasets emphasising faithfulness, evidentiary grounding, and fairness remains an important priority for the field.

 \section*{Acknowledgments}

 The first author acknowledges the support of the Natural Sciences and Engineering Research Council (NSERC) of Canada through the Canada Graduate Scholarships -- Doctoral (CGS D) program. The third author acknowledges the support of the Economic and Social Research Council (ESRC) [grant number ES/P000673/1]. The fourth author acknowledges the support of the Icelandic Research Fund (IRF) [grant number 228511-051]. The last author acknowledges the support of the NSERC [discovery grant RGPIN-2020-07114]. This research was undertaken, in part, thanks to funding from the Canada Research Chairs program [CRC-2024-00192]. This work was enabled in part by support provided by Compute Ontario (\url{https://www.computeontario.ca}), Calcul Qu\'ebec (\url{https://www.calculquebec.ca}), and the Digital Research Alliance of Canada (\url{https://www.alliancecan.ca}).



\section*{Declaration of generative AI and AI-assisted technologies in the manuscript preparation process.}
During the preparation of this work the authors used Claude in order to review text. After using this tool/service, the authors reviewed and edited the content as needed and take full responsibility for the content of the published article.

\section*{Code Availability Statement}

The code and the survey data for this work is available at 
\url{https://github.com/Banking-Analytics-Lab/LLMExplainer}.

\bibliographystyle{elsarticle-harv}
\bibliography{References}

\newpage

\appendix
\counterwithin{table}{section}
\counterwithin{figure}{section}
\renewcommand{\thesection}{Appendix~\Alph{section}}
\renewcommand{\thesubsection}{\Alph{section}.\arabic{subsection}}
\renewcommand\thetable{\Alph{section}.\arabic{table}}
\renewcommand\thefigure{\Alph{section}.\arabic{figure}}

\section{Features used for modelling}
\label{app:Features}

\begin{table*}[hbt!]
\caption{Features used for model training and analysis.}
\label{tab:borrower-features}
\small
\centering
\begin{tabular}{l p{10cm}}
\toprule
Feature & Description \\
\midrule
\texttt{fico} & FICO credit score at origination \\
\texttt{dti} & Original debt-to-income ratio \\
\texttt{ltv} & Original loan-to-value ratio \\
\texttt{mi\_pct} & Mortgage insurance percentage \\
\texttt{orig\_upb} & Original unpaid principal balance \\
\texttt{loan\_term} & Original term of the loan in months \\
\texttt{if\_fthb} & Binary flag indicating whether the borrower is a first-time home buyer \\
\texttt{if\_prim\_res} & Binary flag indicating if the property is the borrower's primary residence \\
\texttt{if\_corr} & Binary indicator for whether the loan originated through a correspondent lender \\
\texttt{if\_sf} & Binary indicator showing whether the property is a single-family home \\
\texttt{if\_purc} & Binary flag indicating whether the loan was used to purchase a home as opposed to refinancing \\
\texttt{cnt\_borr} & Number of borrowers associated with the loan \\
\texttt{cnt\_units} & Number of housing units in the mortgaged property \\
\texttt{default} & Being 90 days or more in payment arrears \\
\bottomrule
\end{tabular}
\end{table*}
\newpage

\section{Mathematical definitions of explainability methods}
\label{app:Mathematical definitions of explainability methods}

This appendix provides formal definitions of SHAP and GNNExplainer, which were summarised in the main text (Section~3.5).

\subsection{SHAP}
Let $f: \mathbb{R}^d \rightarrow \mathbb{R}$ be a predictive model mapping input features $\mathbf{x} = (x_1, \dots, x_d)$ to an output. SHAP decomposes $f(\mathbf{x})$ into contributions from each feature:
\begin{eqnarray}
f(\mathbf{x}) = \phi_0 + \sum_{i=1}^d \phi_i,
\end{eqnarray}
where $\phi_0$ is the expected output, and each $\phi_i$ is a Shapley value:
\begin{eqnarray}
\phi_i = \sum_{S \subseteq N \setminus \{i\}} \frac{|S|!(d - |S| - 1)!}{d!} \Big[ f_{S \cup \{i\}}(\mathbf{x}_{S \cup \{i\}}) - f_S(\mathbf{x}_S) \Big],
\end{eqnarray}
where $N = \{1, 2, \dots, d\}$ is the full set of feature indices and $f_S(\mathbf{x}_S)$ denotes the expected output when only features in subset $S$ are known and the rest are marginalised. Efficient implementations such as TreeSHAP \citep{lundberg2020local} compute these values without exponential cost.

\subsection{GNNExplainer}
For a network $G = (V,E)$ with node features $X \in \mathbb{R}^{|V|\times d}$, GNNExplainer identifies a subgraph $G_S$ and feature mask $M$ that preserve a model’s prediction $f(G,X)_v$ for target node $v$. It maximises the mutual information:
\begin{eqnarray}
\max_{G_S, M} \; \mathbb{I}\big(Y_v; f(G_S, X \odot M)\big),
\end{eqnarray}
where $\odot$ denotes element-wise multiplication, $Y_v$ is the predicted class or score for node $v$, and $\mathbb{I}$ denotes mutual information. This objective is approximated using a differentiable loss:
\begin{eqnarray}
\mathcal{L} = \text{CE}(f(G_S, X \odot M), Y_v) + \lambda_1 \|M\|_1 + \lambda_2 \|A_S\|_1.
\end{eqnarray}
Here, $\text{CE}$ is the cross-entropy loss between the model output and ground truth, $A_S$ is the adjacency matrix of $G_S$, and $\lambda_1,\lambda_2$ are sparsity regularisers.

\newpage

\section{Prompt examples}
\label{app:Prompt examples}

This appendix provides compact, concrete instantiations of the prompt components listed in Table~3 of the main text for each explanation pipeline (tabular-based, network-based, bimodal). Each example shows: (i) predicted outcome, (ii) evidence block(s) passed to the LLM (SHAP and/or GNNExplainer summaries), (iii) structured reasoning instructions and guided subgoals, (iv) counterfactual framing, and (v) explicit output length/format constraints. We omit the shared system prompt for brevity (it is identical across pipelines; see Section~3.8 of the main text).

\paragraph{Shared constraints (applies to all examples)}
The shared system prompt (omitted here) enforces: (a) neutral, evidence-based language; (b) no factors beyond the evidence blocks; (c) no causal guarantees; and (d) explicit length/format constraints. The examples below instantiate these rules.

\begingroup
\footnotesize
\setstretch{0.90}   
\setlength{\tabcolsep}{4pt}
\renewcommand{\arraystretch}{0.90}
\setlist[itemize]{nosep,leftmargin=*,topsep=1pt,partopsep=0pt,itemsep=1pt,parsep=0pt}
\setlist[enumerate]{nosep,leftmargin=*,topsep=1pt,partopsep=0pt,itemsep=1pt,parsep=0pt}

\subsection*{C.1 Tabular-based pipeline prompt example}

\begin{longtable}{L{4.0cm}|L{13.0cm}}
\caption{Tabular-based prompt example instantiating Table 3 of the main text.}
\label{tab:appendix-tabular-prompt}\\
\toprule
Component & Concrete instantiation\\
\midrule
\endfirsthead
\toprule
Component & Concrete instantiation\\
\midrule
\endhead
\bottomrule
\endfoot
1) Predicted outcome &
Prediction: Loan 31780 default probability $=0.8712$. Actual outcome: default.\\
\midrule
2) Top-$k$ SHAP feature attributions with descriptions &
Evidence (top-$k$ SHAP; top-3 shown):
\begin{itemize}
  \item \texttt{cnt\_borr} (SHAP:+0.4362): pushed toward default
  \item \texttt{fico} (SHAP:--0.2609): pushed toward no default
  \item \texttt{dti} (SHAP:+0.0799): pushed toward default
\end{itemize}
(Full top-$k$ list is passed to the LLM; only top-3 shown here.)
Feature descriptions: \texttt{cnt\_borr} = number of borrowers; \texttt{fico} = credit score; \texttt{dti} = debt-to-income ratio.\\
\midrule
3--4) Instructional reasoning + guided subgoals &
Instructions (follow in order):
\begin{enumerate}
  \item Explain the predicted outcome in plain language.
  \item Use the SHAP ordering; explain each listed factor: what it is, whether it increases/decreases risk, and why it matters to lenders.
  \item Combine factors into a short overall justification.
\end{enumerate}
Constraints: borrower-facing language; do not mention SHAP/models/algorithms; reference only evidence-block factors; neutral tone; short sentences.\\
\midrule
5) Borrower-level counterfactual prompt &
Counterfactual guidance: Provide two realistic improvement steps tied to the risk-increasing factors (positive SHAP values). Use cautious phrasing (e.g., ``If you were able to..., this would likely improve...'').\\
\midrule
6) Output format constraint &
Format: No bullet points in the final answer. End with one sentence summarising the main drivers.\\
\end{longtable}

\subsection*{C.2 Network-based pipeline prompt example}

\begin{longtable}{L{4.0cm}|L{13.0cm}}
\caption{Network-based prompt example instantiating Table 3 of the main text.}
\label{tab:appendix-network-prompt}\\
\toprule
Component & Concrete instantiation\\
\midrule
\endfirsthead
\toprule
Component & Concrete instantiation\\
\midrule
\endhead
\bottomrule
\endfoot
1) Predicted outcome &
Prediction: Loan 2885 no-default probability $=0.9489$. Actual outcome: no default.\\
\midrule
2) Network connectivity analysis &
Evidence (GNNExplainer summary; top-3 shown):
\begin{itemize}
  \item Top node features (importance scores):
  \begin{itemize}[leftmargin=1.5em]
    \item \texttt{fico} (0.6967)
    \item \texttt{if\_corr} (0.6736)
    \item \texttt{ltv} (0.3349)
  \end{itemize}
  \item Top edge connections:
  \begin{itemize}[leftmargin=1.5em]
    \item Geographic area connection (Loan 2733 $\rightarrow$ 2885, cluster size $\sim$75): impact 0.2380
    \item Area-provider cluster connection (Loan 2622 $\rightarrow$ 2885, cluster size $\sim$9): impact 0.2314
    \item Same loan provider connection (Loan 2493 $\rightarrow$ 2885, cluster size $\sim$145): impact 0.2266
  \end{itemize}
\end{itemize}
(Full lists are passed to the LLM; only top-3 shown here.)\\
\midrule
3--4) Instructional reasoning + guided subgoals &
Instructions (follow in order):
\begin{enumerate}
  \item Explain the predicted outcome in plain language.
  \item Describe the strongest connections using the edge list and cluster sizes.
  \item Explain (at a high level) why geographic/provider groupings may matter, without adding unverifiable claims.
\end{enumerate}
Constraints: focus on relational/context evidence; mention at least two connection types; do not introduce attributes not shown in the evidence block; neutral tone.\\
\midrule
5) Network-level counterfactual prompt &
Counterfactual guidance: Provide one cautious ``what-if'' statement on how different relational patterns (e.g., fewer links to a given cluster) could change confidence; avoid implying control over other borrowers.\\
\midrule
6) Output format constraint &
Format: Include at least one sentence referencing a connection type and a cluster size from the evidence list. End with a one-sentence summary.\\
\end{longtable}

\subsection*{C.3 Bimodal pipeline prompt example}

\begin{longtable}{L{4.0cm}|L{13.0cm}}
\caption{Bimodal prompt example instantiating Table 3 of the main text.}
\label{tab:appendix-bimodal-prompt}\\
\toprule
Component & Concrete instantiation\\
\midrule
\endfirsthead
\toprule
Component & Concrete instantiation\\
\midrule
\endhead
\bottomrule
\endfoot
1) Predicted outcome &
Prediction: Loan 1007 default probability $=0.9132$. Actual outcome: default.\\
\midrule
2) Top-$k$ SHAP feature descriptions &
Evidence (tabular SHAP; top-3 shown):
\begin{itemize}
  \item \texttt{cnt\_borr} (SHAP=$+0.26$): increases risk
  \item \texttt{dti} (SHAP=$+0.21$): increases risk
  \item \texttt{fico} (SHAP=$-0.18$): reduces risk
\end{itemize}
(Full top-$k$ list is passed to the LLM; only top-3 shown here.)\\
\midrule
3) GNNExplainer subgraph descriptions &
Evidence (network summary; compact subset shown):
\begin{itemize}
  \item Edge: Same provider connection (Loan 5021 $\rightarrow$ 1007, cluster size $\sim$12)
  \item Edge: Area-provider cluster (Loan 2893 $\rightarrow$ 1007, cluster size $\sim$8)
  \item Node: Loan 5021 (defaulted) identified as influential neighbour
\end{itemize}
(Full subgraph/edge lists are passed to the LLM; only a compact subset shown here.)\\
\midrule
4) Combined instructional reasoning linking both modalities &
Instructions (follow in order):
\begin{enumerate}
  \item Explain the predicted outcome overall.
  \item Explain the strongest tabular drivers first (SHAP ordering), then the strongest network drivers (edge types + cluster sizes).
  \item Include one explicit linking sentence describing how borrower-level and relational signals combine.
\end{enumerate}
\\
\midrule
5) Combined borrower- and network-level counterfactual prompt &
Counterfactual guidance: Provide two improvement suggestions: (i) one tied to a positive-SHAP tabular factor; (ii) one stated cautiously about relational context (how different clustering patterns could affect confidence).\\
\midrule
6) Disambiguation hint for conflicting contributions &
Disambiguation rule: If tabular and network evidence point in different directions, acknowledge both and state that the assessment balances the signals, without adding new evidence.\\
\midrule
7) Output format constraint &
Format: Must include (i) at least one tabular feature name from the SHAP list and (ii) at least one connection type plus a cluster size from the network list. End with a one-sentence summary.\\
\end{longtable}
\endgroup

\newpage

\section{Supplementary Methods and Automated Proxy Results}
\label{appendix:D}

This appendix consolidates supplementary methodological detail and secondary automated proxy results moved from the main text to preserve focus on evidence-grounded fidelity and human evaluation. Appendix~D.1 provides full statistical analysis specifications, Appendix~D.2 reports linguistic fluency results via perplexity scoring, and Appendix~D.3 reports simulated stakeholder ratings via Claude~Sonnet~4.6. Results in Appendices~D.2 and~D.3 are treated as complementary proxies to the primary automated and human evaluations reported in Sections~ \ref{Automated evaluation of explanation quality} and \ref{Human evaluation of explanation quality} of the main text.

\subsection{ Statistical Analysis Specifications}
\label{appendix:c1}
Our statistical analysis prioritises robustness and interpretability through effect sizes, uncertainty intervals, and models that account for repeated measures and interaction effects. For human ratings, we use mixed-effects models to account for repeated measures and rater heterogeneity across LLMs, metrics, and evaluators. For automated proxy metrics, we use configuration-level summaries with 95\% confidence intervals and two-way ANOVAs to assess variation between pipelines, LLMs and their interaction. Where interaction effects are significant, the main effects are interpreted cautiously as averages across factor levels, and the cell means are used as the primary basis for more local comparisons.

\paragraph{Automated evaluations (all pipelines)}
For automated evaluations, we report configuration-level means with 95\% confidence intervals for all proxy metrics, including perplexity (linguistic fluency), LLM-as-a-judge fidelity scores (tabular feature coverage, tabular directional consistency, network feature coverage, and network directional consistency), and simulated stakeholder ratings generated separately under CRP- and NCRP-oriented rubrics. To test whether automated scores vary systematically across design choices, we run two-way ANOVA models for each automated metric with factors Pipeline (tabular-based, network-based, bimodal) and LLM (Gemma~3, DeepSeek~R1, Gemini~2.5), including their interaction. We report $F$-statistics, $p$-values, and effect sizes as $\eta^2$, interpreting $\eta^2$ and confidence intervals as primary signals.

\paragraph{Human ratings (bimodal pipeline)}
Human evaluation is limited to bimodal explanations to reduce participant burden. We summarise the mean Likert ratings at the participant-level (1--5) for each metric and LLM (with standard deviations). Although item-level Likert responses are ordinal, previous work supports treating aggregated participant-level means as approximately continuous in mixed designs \citep{Norman2010Likert}. To account for rater-specific leniency/strictness and repeated measures, we fit mixed-effects models with a random intercept for participant and fixed effects for Cohort (CRP vs.\ NCRP), LLM and Metric. We use likelihood-ratio tests to assess whether adding cohort-related terms (and cohort-by-LLM interactions) improves model fit.

\paragraph{Multiplicity control}
Because the human study evaluates eight metrics, we control multiplicity in two complementary ways. For a small set of pre-specified targeted contrasts, we report Holm-corrected paired comparisons, which controls the family-wise error rate (FWER) and ensures that the probability of any single false positive across this confirmatory set remains bounded at $\alpha$. For the broader confirmatory pairwise contrasts within each cohort across the eight metrics, we apply Benjamini-Hochberg False-Discovery-Rate (FDR) control~\citep{benjamini1995controlling}: with the number of simultaneous comparisons well exceeding six, FWER methods become prohibitively conservative~\citep{keselman1999}, and FDR control offers substantially higher power while still bounding the expected proportion of false discoveries.

\paragraph{Reliability and secondary analyses}
To characterise the sources of variation in human ratings, we decompose variance in overall explanation scores using random-intercept models with participant and explanation-case components. Finally, we compare linguistic feature distributions between high- and low-rated explanations and report Cohen's $d$ effect sizes.

\subsection{Fluency via Perplexity Scoring}
\label{appendix:c2}
We assess linguistic fluency using perplexity computed under a frozen GPT-Neo-125M language model trained on the Pile corpus \citep{black2021gptneo, gao2020pile}. Perplexity measures the average surprisal assigned to a sequence, with lower values indicating more predictable, syntactically regular text under the evaluator \citep{holtzman2020curious}. Because our narratives contain domain-specific terminology and numeric information, absolute perplexity may be elevated relative to open-domain text; we therefore interpret perplexity as a relative proxy and compare configurations under a fixed evaluator. Table~\ref{tab:perplexity_scores} reports mean perplexity $\pm$ 95\% confidence intervals.

\begin{table}[hbt!]
\small
\centering
\caption{Perplexity scores across pipeline--LLM configurations.
         Boldface indicates the lowest mean perplexity within each pipeline.}
\label{tab:perplexity_scores}
\resizebox{0.5\textwidth}{!}{%
\begin{tabular}{llc}
\hline
Pipeline & LLM & Perplexity \\
\hline
              & Gemma~3           & 19.38\,$\pm$\,0.67 \\
Tabular-based & DeepSeek~R1       & \textbf{15.93\,$\pm$\,0.36} \\
              & Gemini~2.5        & 33.20\,$\pm$\,1.21 \\
\hline
              & Gemma~3           & 21.09\,$\pm$\,0.59 \\
Network-based & DeepSeek~R1       & \textbf{20.06\,$\pm$\,0.54} \\
              & Gemini~2.5        & 28.96\,$\pm$\,0.72 \\
\hline
        & Gemma~3                & 20.39\,$\pm$\,0.49 \\
Bimodal & DeepSeek~R1            & \textbf{16.77\,$\pm$\,0.43} \\
        & Gemini~2.5             & 37.53\,$\pm$\,5.98 \\
\hline
\end{tabular}%
}
\end{table}

Averaged across LLMs, tabular explanations exhibit the lowest perplexity, network-based explanations are slightly higher, and bimodal explanations are highest, though the ordering depends on the LLM. Across LLMs, DeepSeek~R1 achieves the lowest perplexity within each pipeline, whereas Gemini~2.5 has substantially higher perplexity, particularly in the bimodal pipeline, suggesting that the synthesis of multiple evidence sources can increase the linguistic variability for some LLMs. To test whether perplexity varies systematically across design choices, we fit a two-way ANOVA with the factors Pipeline (tabular-based, network-based, bimodal) and LLM (Gemma~3, DeepSeek~R1, Gemini~2.5), including their interaction (Table~\ref{tab:anova_perplexity}).

\begin{table}[hbt!]
\small
\centering
\renewcommand{\arraystretch}{1.1}
\setlength{\tabcolsep}{6pt}
\begin{tabular}{lccc}
\hline
Effect & $F$ & $p$ & $\eta^2$ \\
\hline
Pipeline & 18.34 & $1.6\times10^{-8}$ & 0.010 \\
LLM & 1302.39 & $<10^{-262}$ & 0.687 \\
Pipeline$\times$LLM & 68.85 & $1.1\times10^{-50}$ & 0.073 \\
\hline
\end{tabular}
\caption{Two-way ANOVA results for perplexity with factors Pipeline and LLM.}
\label{tab:anova_perplexity}
\end{table}

The LLM main effect dominates, explaining the largest share 
of variance in perplexity, while pipeline configuration contributes a smaller but statistically significant effect. The significant interaction coefficient indicates that the relative fluency differences between LLMs depend on the pipeline.

These results should be interpreted with caution. Perplexity is calculated under a fixed and relatively small evaluator, and differences between generators may partly reflect alignment with the evaluator’s training distribution rather than intrinsic writing quality. In addition, variation in explanation length and fine-tuning may contribute to observed differences. We therefore treat perplexity as a comparative indicator of linguistic regularity rather than a definitive measure of explanation quality.

\subsection{Simulated Ratings via Claude~Sonnet~4.6}
\label{appendix:c3}
To approximate audience-conditioned perceptions of interpretability and decision support, we collect role-conditioned simulated Likert ratings using Claude~Sonnet~4.6 \citep{anthropic2024claude} to approximate CRP- and NCRP-oriented perceptions. The judge LLM was prompted to rate each explanation from two perspectives: a CRP viewpoint that emphasises evidence and operational relevance, and an NCRP viewpoint that prioritises clarity and perceived helpfulness. Tables~\ref{tab:crp_merged_matrix} and~\ref{tab:ncrp_merged_matrix} report simulated CRP and NCRP ratings in all pipeline--LLM configurations. Values are reported as mean $\pm$ 95\% confidence intervals.

\begin{table}[hbt!]
\small
\centering
\caption{Simulated CRP ratings across pipeline--LLM configurations.
         Boldface indicates the highest mean per column.}
\label{tab:crp_merged_matrix}
\resizebox{\textwidth}{!}{%
\begin{tabular}{llcccccccc}
\hline
Pipeline & LLM & UND & TRU & INS & SAT & CON & CVN & COM & USB \\
\hline
              & Gemma~3           & 3.89\,$\pm$\,0.07 & 2.11\,$\pm$\,0.06 & 2.15\,$\pm$\,0.07 & 1.99\,$\pm$\,0.02 & 2.62\,$\pm$\,0.10 & 2.11\,$\pm$\,0.06 & 2.70\,$\pm$\,0.09 & 1.42\,$\pm$\,0.10 \\
Tabular-based & DeepSeek~R1       & 3.79\,$\pm$\,0.08 & 2.38\,$\pm$\,0.10 & 2.34\,$\pm$\,0.09 & 2.14\,$\pm$\,0.07 & 2.75\,$\pm$\,0.09 & 2.38\,$\pm$\,0.10 & 2.95\,$\pm$\,0.07 & 1.79\,$\pm$\,0.08 \\
              & Gemini~2.5        & \textbf{3.99\,$\pm$\,0.08} & \textbf{2.80\,$\pm$\,0.08} & \textbf{2.78\,$\pm$\,0.09} & \textbf{2.58\,$\pm$\,0.10} & \textbf{2.96\,$\pm$\,0.08} & \textbf{2.79\,$\pm$\,0.08} & \textbf{3.29\,$\pm$\,0.12} & \textbf{1.86\,$\pm$\,0.07} \\
\hline
               & Gemma~3          & 3.62\,$\pm$\,0.10 & 1.97\,$\pm$\,0.03 & 1.98\,$\pm$\,0.03 & 1.96\,$\pm$\,0.04 & 2.15\,$\pm$\,0.08 & 1.97\,$\pm$\,0.03 & 2.78\,$\pm$\,0.08 & 1.12\,$\pm$\,0.06 \\
Network-based  & DeepSeek~R1      & 2.98\,$\pm$\,0.03 & 1.77\,$\pm$\,0.08 & 1.81\,$\pm$\,0.08 & 1.57\,$\pm$\,0.10 & 1.91\,$\pm$\,0.06 & 1.77\,$\pm$\,0.08 & 2.02\,$\pm$\,0.04 & 1.00\,$\pm$\,0.00 \\
               & Gemini~2.5       & 2.80\,$\pm$\,0.08 & 1.40\,$\pm$\,0.10 & 1.55\,$\pm$\,0.10 & 1.29\,$\pm$\,0.09 & 1.78\,$\pm$\,0.08 & 1.41\,$\pm$\,0.10 & 1.90\,$\pm$\,0.07 & 1.00\,$\pm$\,0.00 \\
\hline
         & Gemma~3               & 3.94\,$\pm$\,0.05 & 2.51\,$\pm$\,0.10 & 2.49\,$\pm$\,0.10 & 2.01\,$\pm$\,0.04 & 2.80\,$\pm$\,0.08 & 2.49\,$\pm$\,0.10 & 2.73\,$\pm$\,0.09 & 1.62\,$\pm$\,0.10 \\
Bimodal  & DeepSeek~R1           & 3.98\,$\pm$\,0.03 & 2.51\,$\pm$\,0.10 & 2.58\,$\pm$\,0.10 & 2.24\,$\pm$\,0.09 & 2.94\,$\pm$\,0.05 & 2.50\,$\pm$\,0.10 & 3.00\,$\pm$\,0.08 & 1.83\,$\pm$\,0.07 \\
         & Gemini~2.5            & 3.85\,$\pm$\,0.08 & 2.48\,$\pm$\,0.11 & 2.59\,$\pm$\,0.10 & 2.19\,$\pm$\,0.08 & 2.75\,$\pm$\,0.09 & 2.47\,$\pm$\,0.11 & 2.62\,$\pm$\,0.11 & 1.62\,$\pm$\,0.10 \\
\hline
\end{tabular}%
}
\end{table}

\begin{table}[hbt!]
\small
\centering
\caption{Simulated NCRP ratings across pipeline--LLM configurations.
         Boldface indicates the highest mean per column.}
\label{tab:ncrp_merged_matrix}
\resizebox{\textwidth}{!}{%
\begin{tabular}{llcccccccc}
\hline
Pipeline & LLM & UND & TRU & INS & SAT & CON & CVN & COM & USB \\
\hline
              & Gemma~3           & \textbf{4.43\,$\pm$\,0.11} & 3.44\,$\pm$\,0.10 & 3.89\,$\pm$\,0.06 & 3.45\,$\pm$\,0.10 & 3.43\,$\pm$\,0.10 & 4.13\,$\pm$\,0.10 & \textbf{4.43\,$\pm$\,0.11} & 3.38\,$\pm$\,0.10 \\
Tabular-based & DeepSeek~R1       & 4.00\,$\pm$\,0.08 & 3.20\,$\pm$\,0.10 & 3.85\,$\pm$\,0.08 & 3.36\,$\pm$\,0.10 & 3.19\,$\pm$\,0.10 & 3.91\,$\pm$\,0.08 & 4.00\,$\pm$\,0.08 & 3.10\,$\pm$\,0.09 \\
              & Gemini~2.5        & 4.39\,$\pm$\,0.10 & \textbf{3.62\,$\pm$\,0.10} & \textbf{4.19\,$\pm$\,0.08} & 3.81\,$\pm$\,0.08 & \textbf{3.62\,$\pm$\,0.10} & \textbf{4.29\,$\pm$\,0.11} & 4.39\,$\pm$\,0.10 & \textbf{3.54\,$\pm$\,0.10} \\
\hline
               & Gemma~3          & 3.97\,$\pm$\,0.07 & 2.97\,$\pm$\,0.07 & 3.68\,$\pm$\,0.11 & 3.02\,$\pm$\,0.06 & 2.97\,$\pm$\,0.07 & 3.86\,$\pm$\,0.07 & 3.97\,$\pm$\,0.07 & 2.97\,$\pm$\,0.07 \\
Network-based  & DeepSeek~R1      & 3.63\,$\pm$\,0.10 & 2.64\,$\pm$\,0.10 & 2.91\,$\pm$\,0.06 & 2.65\,$\pm$\,0.10 & 2.64\,$\pm$\,0.10 & 3.02\,$\pm$\,0.06 & 3.63\,$\pm$\,0.10 & 2.41\,$\pm$\,0.10 \\
               & Gemini~2.5       & 3.09\,$\pm$\,0.08 & 2.04\,$\pm$\,0.06 & 2.45\,$\pm$\,0.10 & 2.04\,$\pm$\,0.06 & 2.04\,$\pm$\,0.06 & 2.93\,$\pm$\,0.06 & 3.09\,$\pm$\,0.08 & 2.02\,$\pm$\,0.06 \\
\hline
         & Gemma~3               & 4.12\,$\pm$\,0.07 & 3.21\,$\pm$\,0.09 & 3.99\,$\pm$\,0.02 & 3.63\,$\pm$\,0.10 & 3.19\,$\pm$\,0.08 & 3.99\,$\pm$\,0.04 & 4.13\,$\pm$\,0.07 & 3.19\,$\pm$\,0.08 \\
Bimodal  & DeepSeek~R1           & 4.15\,$\pm$\,0.07 & 3.18\,$\pm$\,0.08 & 4.00\,$\pm$\,0.00 & 3.47\,$\pm$\,0.10 & 3.17\,$\pm$\,0.07 & 4.00\,$\pm$\,0.00 & 4.15\,$\pm$\,0.07 & 3.17\,$\pm$\,0.07 \\
         & Gemini~2.5            & 4.40\,$\pm$\,0.12 & 3.40\,$\pm$\,0.11 & 4.07\,$\pm$\,0.09 & \textbf{3.82\,$\pm$\,0.09} & 3.39\,$\pm$\,0.11 & 4.09\,$\pm$\,0.08 & 4.40\,$\pm$\,0.12 & 3.38\,$\pm$\,0.11 \\
\hline
\end{tabular}%
}
\end{table}

The simulated ratings exhibit three descriptive trends. First, NCRP-oriented ratings are consistently higher than CRP-oriented ratings across metrics, suggesting that explanations optimised for clarity and surface fluency are more positively received under the NCRP rubric than under the evidence- and compliance-oriented CRP rubric.

Second, across CRP-oriented ratings, tabular-based and bimodal pipelines generally outperform network-based pipelines, indicating that explanations grounded in borrower-level evidence, whether alone or combined with relational signals, are rated more favourably than network-only explanations. Within this pattern, Gemini~2.5 often achieves the highest mean CRP ratings, although the differences between tabular-based and bimodal pipelines are modest and metric-dependent.

Third, network-based pipelines consistently receive the lowest CRP ratings across LLMs, indicating that abstract relational evidence is comparatively difficult to operationalise in ways that align with evidence-driven evaluation criteria. In contrast, NCRP ratings show stronger sensitivity to communicative quality, with large pipeline effects and additional (though smaller) LLM and interaction effects concentrated on communicative dimensions (e.g., UND, CVN, COM). To assess whether these descriptive patterns reflect systematic variation between design choices, we fit two-way ANOVAs separately for simulated CRP and NCRP ratings for each metric, with factors Pipeline (tabular-based, network-based, bimodal) and LLM (Gemma~3, DeepSeek~R1, Gemini~2.5), including their interaction. We report $F$-statistics, $p$-values, and effect sizes ($\eta^2$) in Tables~\ref{tab:anova_simulated_crp} and~\ref{tab:anova_simulated_ncrp}. The corresponding cell means in Tables~\ref{tab:crp_merged_matrix}--\ref{tab:ncrp_merged_matrix} provide the main basis for within-pipeline comparisons.

\begin{table}[hbt!]
\small
\centering
\setlength{\tabcolsep}{6pt}
\renewcommand{\arraystretch}{0.9}
\caption{Two-way ANOVA results for simulated CRP ratings.}
\label{tab:anova_simulated_crp}
\resizebox{\columnwidth}{!}{%
\begin{tabular}{lccccccccc}
\toprule
Metric
& $F_{\text{pipe}}$ & $p_{\text{pipe}}$ & $\eta^2_{\text{pipe}}$
& $F_{\text{llm}}$  & $p_{\text{llm}}$  & $\eta^2_{\text{llm}}$
& $F_{\text{int}}$  & $p_{\text{int}}$  & $\eta^2_{\text{int}}$ \\
\midrule
UND & 471.115 & $8.97\times10^{-140}$ & 0.517 & 51.290 & $8.55\times10^{-22}$ & 0.105 & 54.234 & $6.95\times10^{-41}$ & 0.198 \\
TRU & 288.654 & $4.30\times10^{-97}$  & 0.396 & 0.575  & 0.5628 & 0.001 & 52.728 & $7.66\times10^{-40}$ & 0.194 \\
INS & 260.980 & $1.07\times10^{-89}$  & 0.373 & 3.884  & 0.0209 & 0.009 & 38.091 & $2.03\times10^{-29}$ & 0.148 \\
SAT & 243.487 & $7.18\times10^{-85}$  & 0.357 & 0.930  & 0.3951 & 0.002 & 78.371 & $7.06\times10^{-57}$ & 0.263 \\
CON & 445.581 & $2.41\times10^{-134}$ & 0.503 & 0.512  & 0.5996 & 0.001 & 22.268 & $1.56\times10^{-17}$ & 0.092 \\
CVN & 279.329 & $1.24\times10^{-94}$  & 0.389 & 0.574  & 0.5632 & 0.001 & 50.728 & $1.89\times10^{-38}$ & 0.188 \\
COM & 229.206 & $7.76\times10^{-81}$  & 0.343 & 6.907  & 0.0011 & 0.015 & 88.401 & $3.72\times10^{-63}$ & 0.287 \\
USB & 294.177 & $1.55\times10^{-98}$  & 0.401 & 13.379 & $1.89\times10^{-6}$ & 0.030 & 20.158 & $6.71\times10^{-16}$ & 0.084 \\
\bottomrule
\end{tabular}
}
\end{table}

\begin{table}[hbt!]
\small
\centering
\setlength{\tabcolsep}{6pt}
\renewcommand{\arraystretch}{0.9}
\caption{Two-way ANOVA results for simulated NCRP ratings.}
\label{tab:anova_simulated_ncrp}
\resizebox{\columnwidth}{!}{%
\begin{tabular}{lccccccccc}
\toprule
Metric
& $F_{\text{pipe}}$ & $p_{\text{pipe}}$ & $\eta^2_{\text{pipe}}$
& $F_{\text{llm}}$  & $p_{\text{llm}}$  & $\eta^2_{\text{llm}}$
& $F_{\text{int}}$  & $p_{\text{int}}$  & $\eta^2_{\text{int}}$ \\
\midrule
UND & 229.281 & $7.38\times10^{-81}$  & 0.343 & 27.297 & $3.15\times10^{-12}$ & 0.058 & 53.932 & $1.12\times10^{-40}$ & 0.197 \\
TRU & 302.394 & $1.16\times10^{-100}$ & 0.408 & 17.788 & $2.67\times10^{-8}$  & 0.039 & 57.377 & $4.83\times10^{-43}$ & 0.207 \\
INS & 685.194 & $4.47\times10^{-180}$ & 0.609 & 54.104 & $6.93\times10^{-23}$ & 0.110 & 123.595 & $1.06\times10^{-83}$ & 0.260 \\
SAT & 503.172 & $2.23\times10^{-146}$ & 0.534 & 16.600 & $8.39\times10^{-8}$  & 0.036 & 72.315 & $5.50\times10^{-53}$ & 0.248 \\
CON & 297.011 & $2.85\times10^{-99}$  & 0.403 & 16.985 & $5.79\times10^{-8}$  & 0.037 & 58.700 & $6.06\times10^{-44}$ & 0.211 \\
CVN & 461.246 & $1.08\times10^{-137}$ & 0.512 & 68.807 & $1.73\times10^{-28}$ & 0.135 & 75.891 & $2.71\times10^{-55}$ & 0.257 \\
COM & 229.347 & $7.07\times10^{-81}$  & 0.343 & 27.964 & $1.68\times10^{-12}$ & 0.060 & 53.075 & $4.40\times10^{-40}$ & 0.195 \\
USB & 332.705 & $2.64\times10^{-108}$ & 0.431 & 31.776 & $4.74\times10^{-14}$ & 0.067 & 55.375 & $1.14\times10^{-41}$ & 0.201 \\
\bottomrule
\end{tabular}
}
\end{table}

In general, the ANOVA results show substantial systematic variation in simulated ratings between design choices. For both CRP- and NCRP-oriented ratings, pipeline effects are large, indicating that evidence modality is the main driver of perceived quality, while Pipeline$\times$LLM interactions show that LLM performance depends on the evidence provided. The LLM main effects are smaller overall: under the CRP rubric they are limited and most visible for clarity-related metrics such as UND (and, to a lesser extent, USB), whereas under the NCRP rubric they are more pronounced on communicative dimensions such as UND, COM, and USB.

Nevertheless, because Claude~Sonnet~4.6 is itself an LLM with its own stylistic tendencies, independence is relative rather than absolute, and these simulated ratings may reflect judge LLM preferences in addition to the intended audience-conditioned rubric. We therefore treat simulated ratings as complementary proxies and prioritise the human survey results as the primary reference standard.

\newpage

\section{Representative Explanation Examples}
\label{appendix:E}

This appendix provides representative explanation examples from all three pipelines: tabular-based (E.1), network-based (E.2), and bimodal (E.3), the last of which was discussed in Section \ref{Descriptive overview of representative explanations} of the main text.

\subsection{Tabular-Based Pipeline Example}
\label{appendix:d1}
Explanations emphasise loan-level attributes such as credit score, debt-to-income ratio, and loan-to-value. They typically adopt a deterministic framing and highlight the top-ranked features. Although clear and concise, these outputs may oversimplify trade-offs in borderline cases.

\begin{exampleframe}
\textbf{Example from Gemma~3:}
\begin{textit}
Your loan application has a 97\% chance of being approved, which is much better than most people. This is largely because your credit score of 744 is excellent. A high credit score shows banks that you're reliable when it comes to paying back money, making them more confident in lending to you.

Another big plus is your monthly debts, which are lower than most borrowers. This tells banks that you have room in your budget to handle new payments, reducing the risk for them. Applying for the loan by yourself is also a positive factor because it simplifies the decision-making process for the bank. Plus, choosing a shorter loan term is smart because it means you'll pay less interest over time, which banks like because it shows you're focused on saving money. 

All these factors together make you a strong candidate for approval. Overall, the strongest drivers in this case are the credit score and the relatively low level of monthly debt.
\end{textit}
\end{exampleframe}

\subsection{Network-Based Pipeline Example}
\label{appendix:d2}
Explanations highlight how a loan’s relational structure (e.g., provider or geographic clusters) shapes predicted risk. Narratives often reference neighbouring loans or cluster-level patterns as supporting evidence, thereby adding relational context (i.e., group membership effects). However, the resulting explanations can be more abstract than feature-based ones.

\begin{exampleframe}
\textbf{Example from DeepSeek~R1:}
\begin{textit}
Your loan application has a 68\% chance of being approved, which is a promising indicator. The main reason for this positive outlook is your network connections in your area and with your lender.

You are connected to 2 other borrowers in your geographic area, and these connections show positive patterns. This suggests that your area has a stable job market and strong housing values, which are important factors for lenders. Lenders like to see that borrowers are in areas with steady employment opportunities and consistent property values, as these factors reduce the risk of lending.

Additionally, you share the same lender type with 1 other borrower, and this also shows positive patterns. This indicates that the lender has experience lending in your area and understands the local economic conditions, which can make them more confident in approving loans here.

You are also connected to 1 other borrower who both lives in your area and uses the same lender type, further reinforcing these positive patterns. These connections to a total of 4 borrowers help your approval chances by demonstrating a pattern of successful lending in your area.

To maintain or strengthen these network advantages, consider staying informed about your local job market and housing trends. Compared to other borrowers in similar network situations, your connections are quite strong, which works in your favour.

Overall, your network connections suggest a stable and supportive local economic environment, which is beneficial for your loan application. By understanding and leveraging these connections, you can continue to strengthen your position as a borrower.
\end{textit}
\end{exampleframe}

\subsection{Bimodal Pipeline Example}
Bimodal explanations integrate tabular and relational evidence into a single narrative. When both signal types align, this yields more holistic rationales; when they conflict, careful structuring is required to avoid contradictions. The following example illustrates this integration.

\begin{exampleframe}
\textbf{Example from Gemini~2.5:}
\begin{textit}
Based on our thorough review, your loan application has a strong 75\% chance of approval, which is a very positive outlook.

First, your excellent credit score of 809 truly stands out, showing a very responsible history of managing money, much better than most borrowers. While you're applying solo, meaning we're looking solely at your financial strength, your debt-to-income ratio is a bit higher than 40\% of other applicants, indicating more of your income goes towards existing monthly payments. However, you're also putting down a significant down payment – more than 80\% of people – and requesting a much smaller loan amount than most, making it very manageable.

Beyond your personal finances, your local connections are also a positive sign; we see you're connected to 3 other borrowers in your geographic area and 2 more who also use our lender type, totaling 5 positive connections that show stable patterns in your community. These combined personal strengths, like your strong credit and down payment, along with positive local patterns, significantly contribute to your high approval likelihood. To make your application even stronger, consider focusing on paying down credit cards or other debts over the next 6-12 months; this would significantly improve your debt-to-income ratio, making your financial picture even more appealing. 

Overall, your complete picture is very strong, demonstrating a solid financial foundation and positive community ties, which gives us confidence in this assessment.
\end{textit}
\end{exampleframe}

\newpage

\section{Participant demographics and additional survey information}
\label{app:demographics}

This appendix reports supplementary statistics about the survey population used in Section \ref{Human evaluation of explanation quality} of the main text. Differences in the numbers of CRPs and NCRPs across the figures are due to the fact that the demographic questions in the survey were optional.

\begin{figure*}[hbt!]
\centering
\begin{subfigure}[t]{0.44\textwidth}
    \centering
    \includegraphics[width=\linewidth]{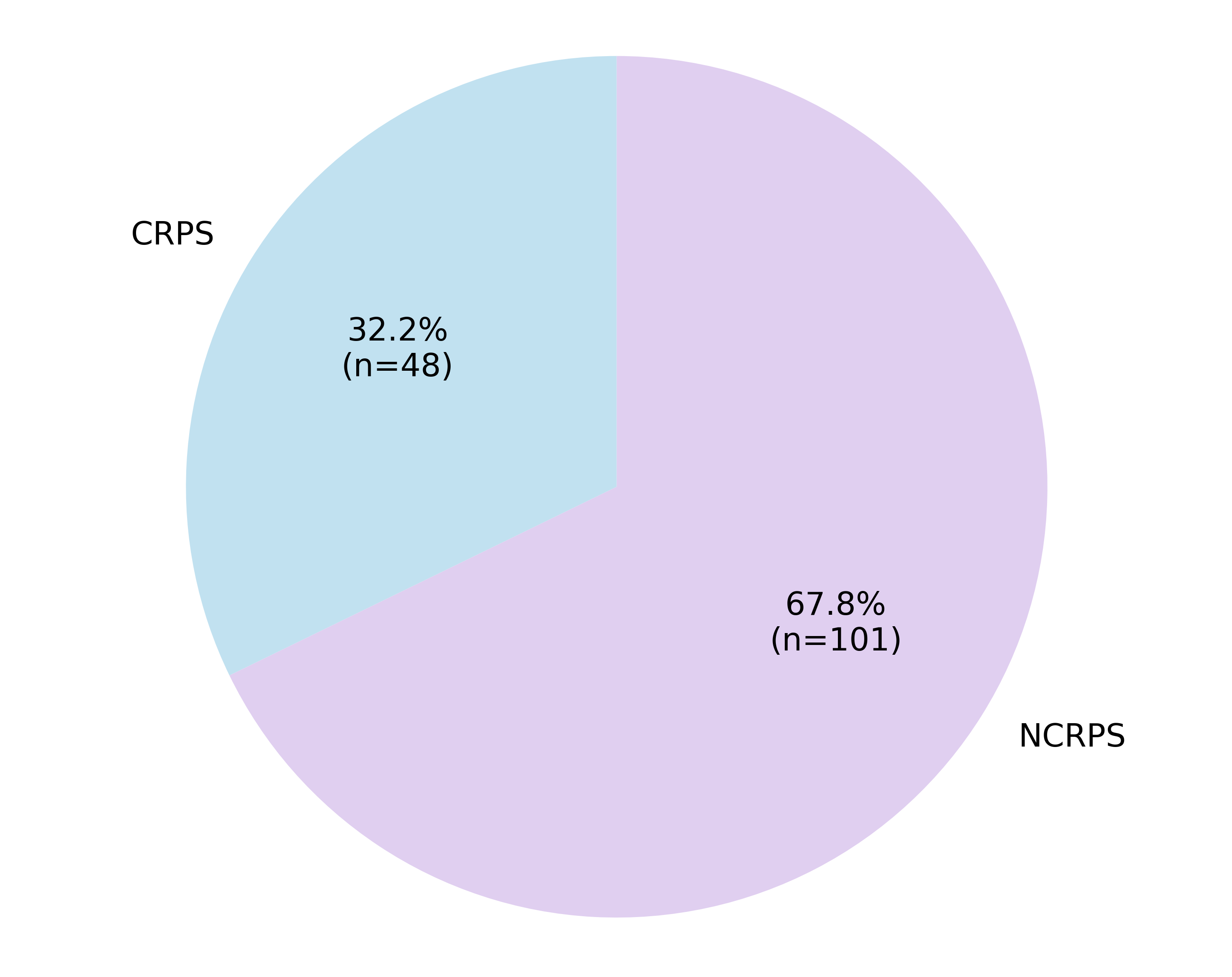}
    \caption{Cohort composition}
    \label{fig:cohort}
\end{subfigure}\hfill
\begin{subfigure}[t]{0.48\textwidth}
    \centering
    \includegraphics[width=\linewidth]{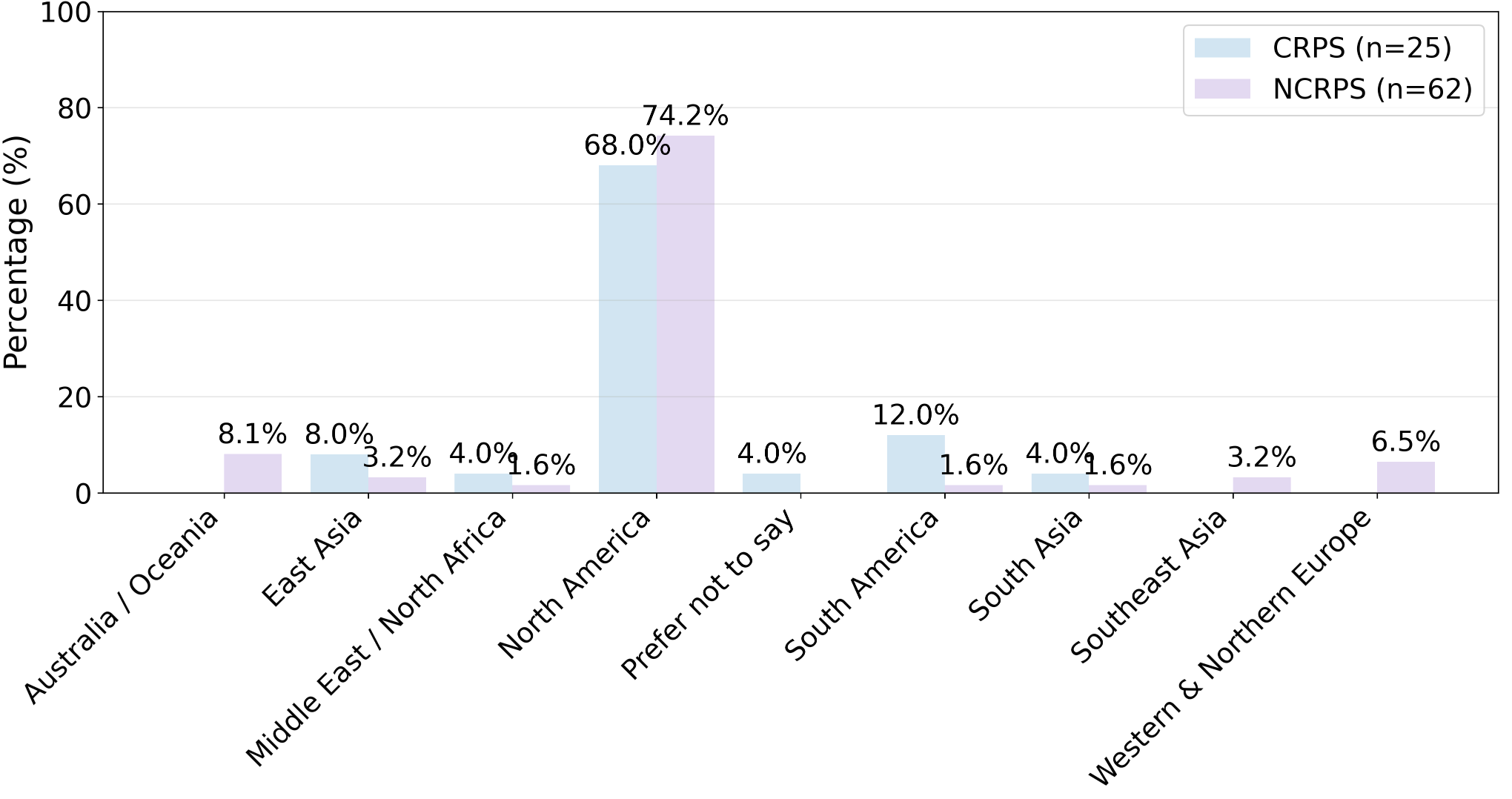}
    \caption{Geographic regions}
    \label{fig:geo}
\end{subfigure}

\vspace{1em}

\begin{subfigure}[t]{0.48\textwidth}
    \centering
    \includegraphics[width=\linewidth]{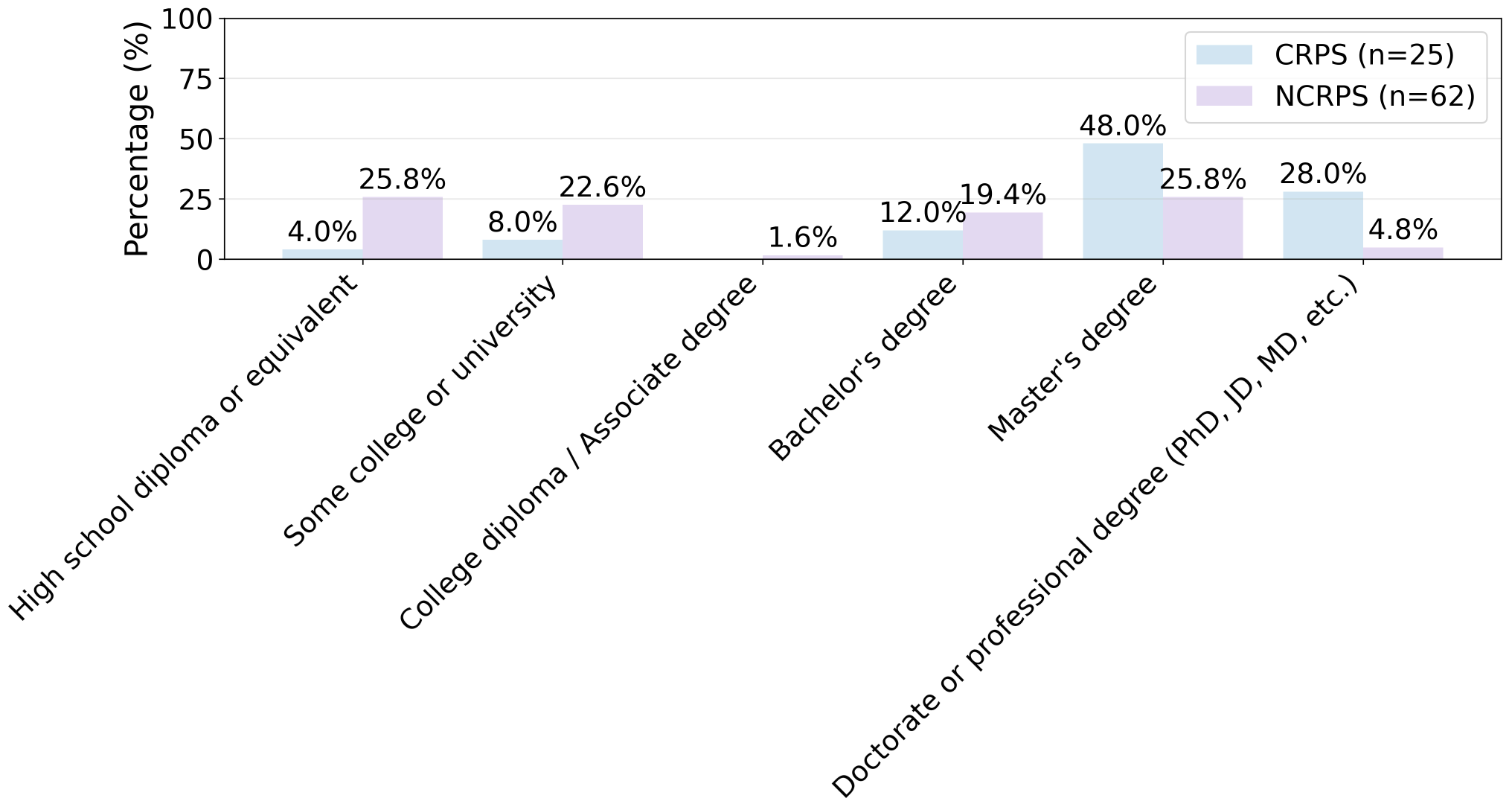}
    \caption{Educational background}
    \label{fig:education}
\end{subfigure}\hfill
\begin{subfigure}[t]{0.48\textwidth}
    \centering
    \includegraphics[width=\linewidth]{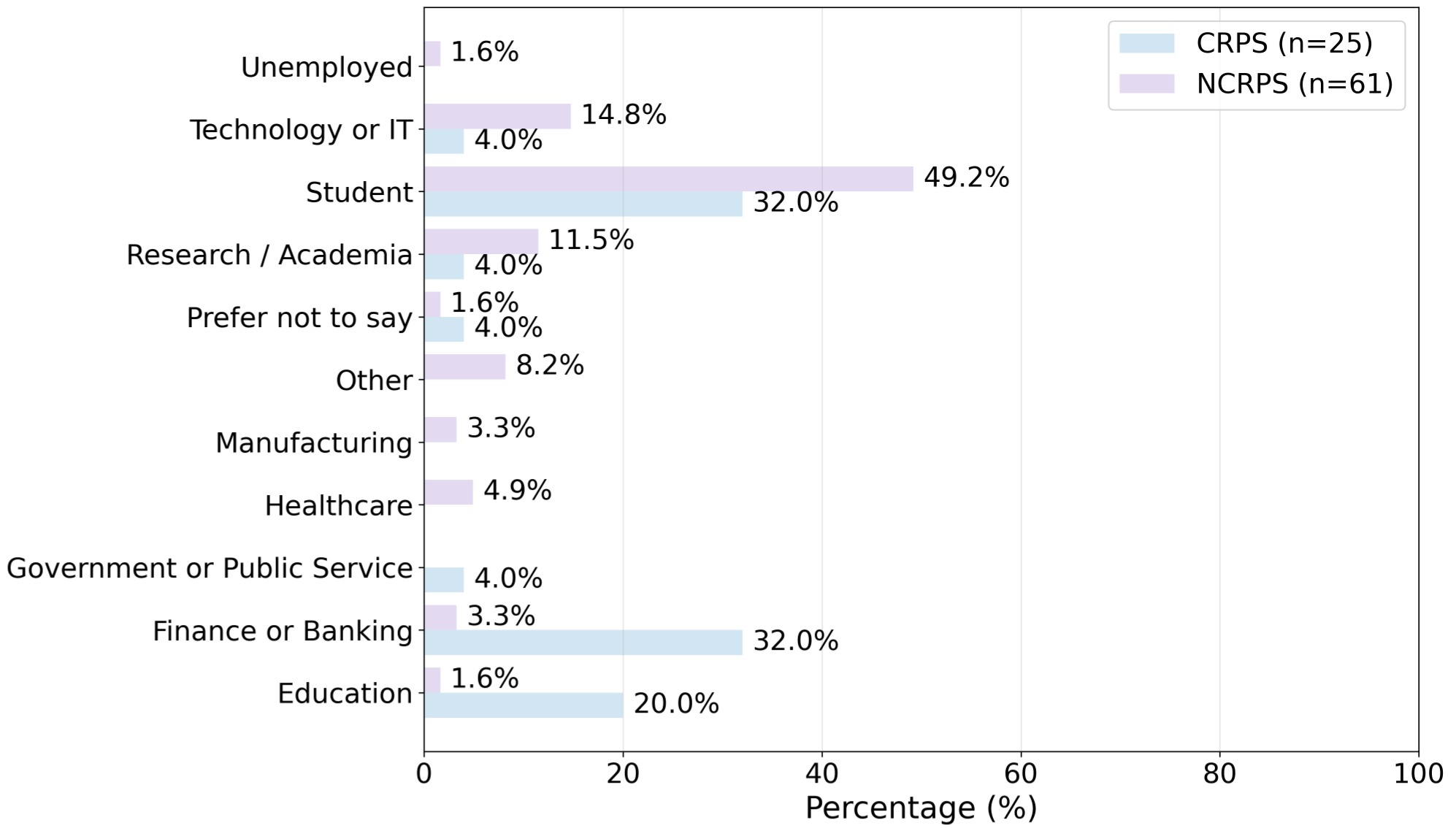}
    \caption{Employment/economic sector}
    \label{fig:sector}
\end{subfigure}

\vspace{1em}

\begin{subfigure}[t]{0.48\textwidth}
    \centering
    \includegraphics[width=\linewidth]{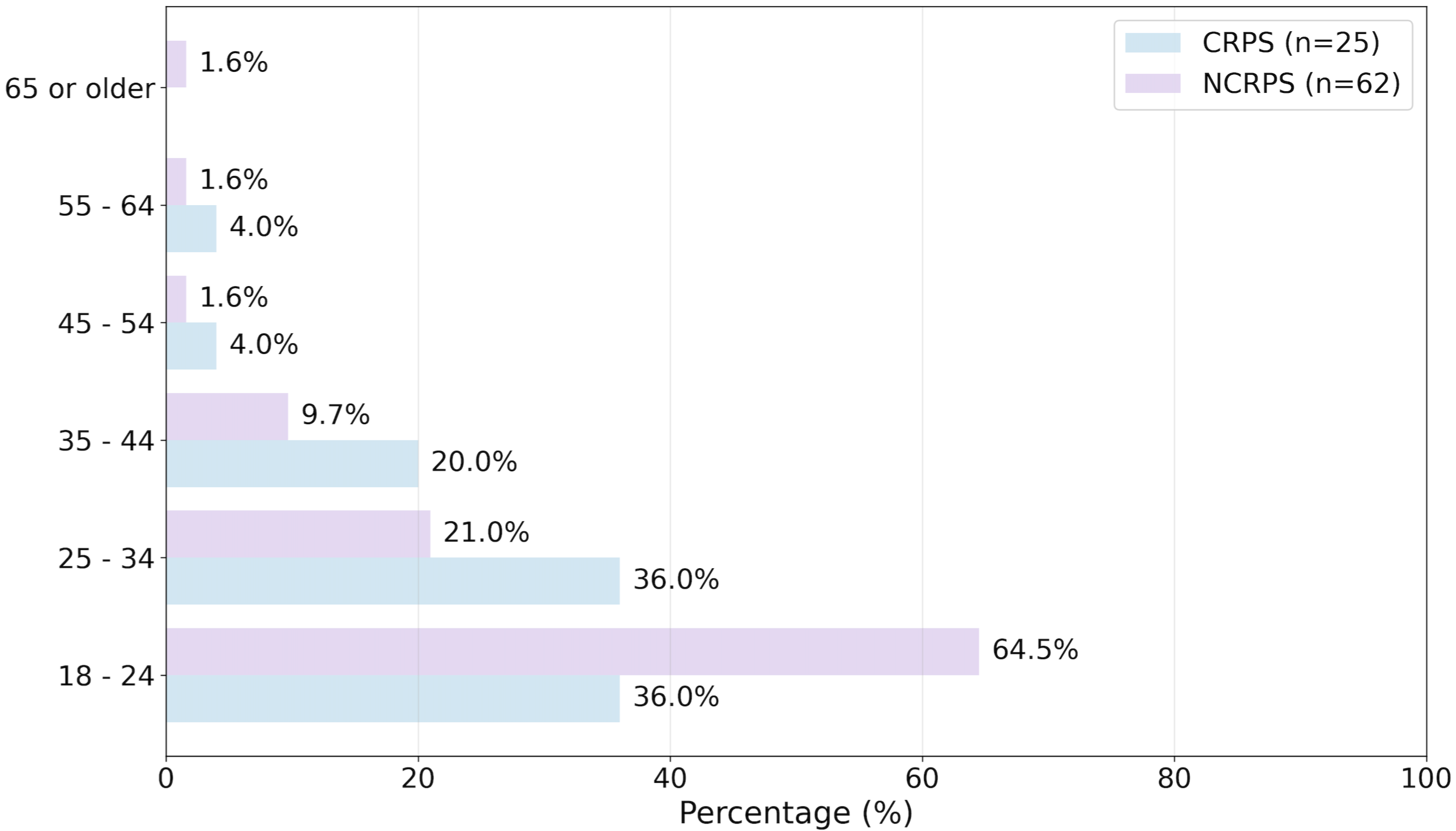}
    \caption{Age distribution}
    \label{fig:age}
\end{subfigure}\hfill
\begin{subfigure}[t]{0.48\textwidth}
    \centering
    \includegraphics[width=\linewidth]{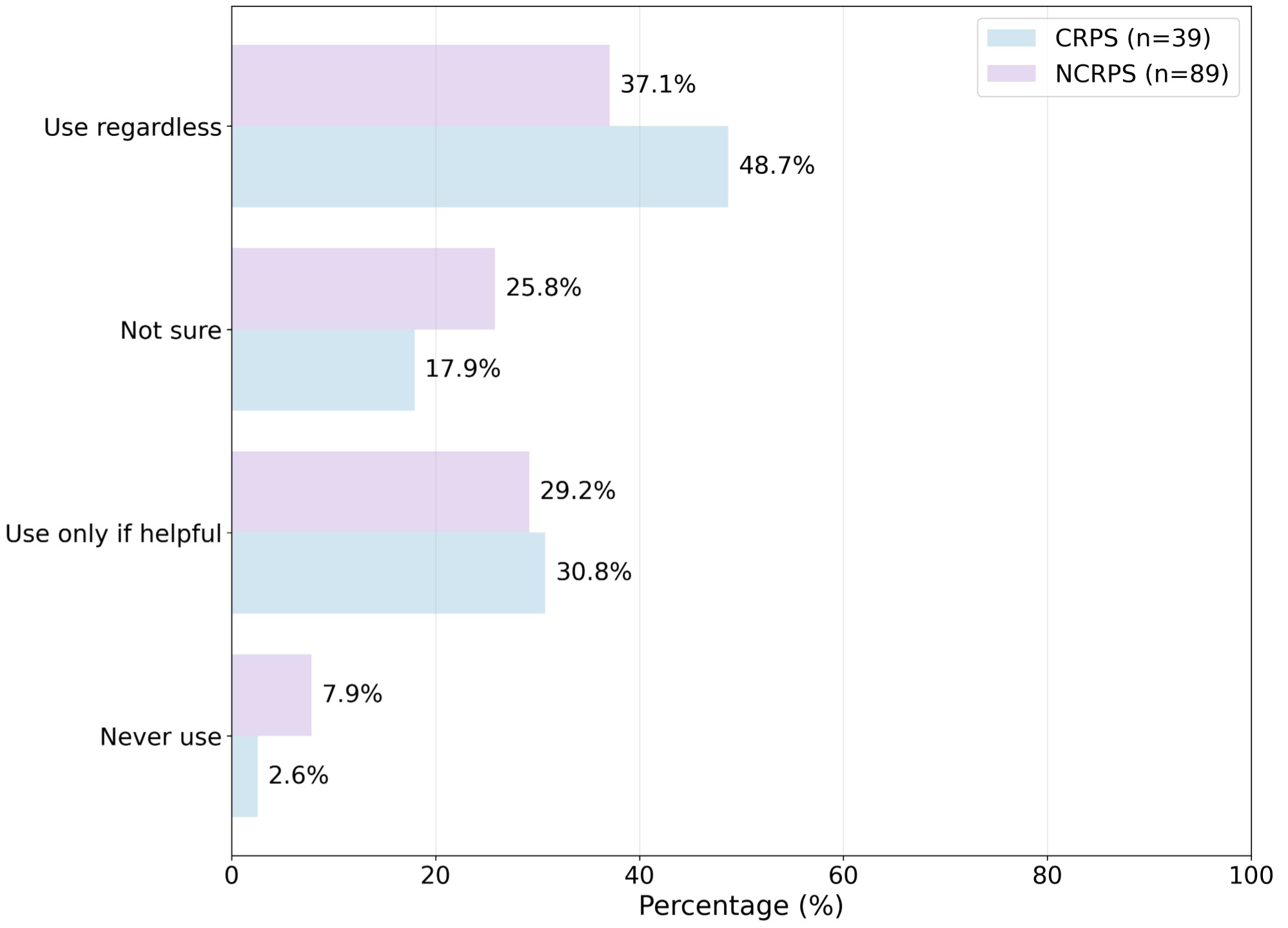}
    \caption{Attitudes toward network variables}
    \label{fig:credit}
\end{subfigure}

\caption{Demographic and background characteristics of survey participants.}
\label{fig:demographics}
\end{figure*}

\end{document}